\documentclass[12pt,cls,onecolumn]{IEEEtran}
\usepackage{graphicx}
\usepackage{amssymb}
\usepackage{amsmath}
\usepackage{cite}
\usepackage{cases}
\usepackage{color}
\usepackage{epstopdf}
\usepackage{dsfont}
\usepackage{mdwlist}
\usepackage{bigstrut,array,multirow,tabularx}
\usepackage{dsfont}
\usepackage{bbm}
\usepackage{algorithm,algorithmic}

\renewcommand{\IEEEQED}{\IEEEQEDopen}
\AtBeginDocument{}

\DeclareMathAlphabet{\mathsf}{OML}{cmbr}{m}{it}

\newtheorem{theorem}{Theorem}
\newtheorem{lemma}{Lemma}

\newtheorem{proposition}{Proposition}

\newtheorem{remark}{Remark}

\DeclareMathOperator{\tr}{\mathrm{tr}}
\DeclareMathOperator{\E}{\mathds{E}}

\newcommand{\C}{\mathbbmss{C}}
\newcommand{\B}[1]{\mathbf{#1}}

\newcommand{\RV}[1]{\mathsf{#1}}
\newcommand{\EX}[1]{\E\left\{{#1}\right\}}

\newcommand{\IM}[1]{\B{I}_{#1}}

\newcommand{\rank}[1]{\textsf{rank}\left(#1\right)}
\newcommand{\diag}[1]{\mathrm{diag}\left(#1\right)}

\newcommand{\BB}[1]{{\pmb{#1}}}
\newcommand{\cvH}[1]{\BB{h}_{#1}}
\newcommand{\cvG}[1]{\BB{g}_{#1}}
\newcommand{\he}{\BB{h}_{\mathrm{e}}}

\newcommand{\rbf}{\BB{v}}
\newcommand{\tbf}{\BB{w}}

\newcommand{\zv}{\BB{z}}
\newcommand{\uv}{\BB{u}}

\newcommand{\wv}{\BB{w}}
\newcommand{\nv}{\BB{n}}
\newcommand{\fv}{\BB{f}}
\newcommand{\ch}{\BB{H}}

\newcommand{\G}{\BB{G}}
\newcommand{\Y}{\BB{Y}}
\newcommand{\Z}{\BB{Z}}
\newcommand{\W}{\BB{W}}
\newcommand{\D}{\B{D}}
\newcommand{\Q}{\BB{Q}}
\newcommand{\nQ}{\bar{\BB{Q}}}
\newcommand{\nW}{\bar{\BB{W}}}
\newcommand{\F}{\BB{F}}
\newcommand{\nSx}{n_{\mathrm{s}}}
\newcommand{\nRx}{n_{\mathrm{r}}}
\newcommand{\nBx}{n_{\mathrm{b}}}
\newcommand{\Svar}{\delta_{\theta}}
\newcommand{\card}[1]{\left[#1\right]}
\newcommand{\PL}[1]{\textsf{PL}\left(#1\right)}
\newcommand{\av}{\BB{a}}
\newcommand{\ampl}{\alpha}
\newcommand{\Ampl}{\B{A}}
\newcommand{\Rs}{\BB{R}_{\mathrm{s}}}
\newcommand{\Rn}{\BB{R}_{\mathrm{n}}}
\newcommand{\RNvar}[1]{\sigma_{\mathrm{n},#1}}
\newcommand{\nvar}{\sigma_{\mathrm{n}}}
\newcommand{\RFCvar}{\sigma_{\mathrm{tot}}}
\newcommand{\Nvar}[1]{\sigma_{\mathrm{u},#1}}
\newcommand{\Pf}[2]{\Phi_{#1}\left(#2\right)}

\newcommand{\maxeig}{\lambda_{\mathrm{max}}}

\newcommand{\f}[2]{f_{#1}\left(#2\right)}

\newcommand{\mE}[3]{\left(#1\right)_{#2,#3}}
\newcommand{\Norm}[1]{\left|{#1}\right|}
\newcommand{\Vnorm}[1]{\left\|{#1}\right\|}

\newcommand{\Pb}[1]{(\textsf{P#1})}
\newcommand{\sdr}[1]{(\textsf{SDR#1})}
\newcommand{\PbS}[1]{(\textsf{P#1--Sum})}

\begin{document}

\title{Wireless Power Transfer for Distributed Estimation in Sensor Networks}
\author{\large Vien V. Mai, Won-Yong Shin, \emph{Senior Member}, \emph{IEEE}, \\ and Koji Ishibashi, \emph{Member}, \emph{IEEE}
\\
\thanks{This research was supported by the Basic Science Research Program
through the National Research Foundation of Korea (NRF) funded by
the Ministry of Education (2014R1A1A2054577) and by the Ministry
of Science, ICT \& Future Planning (MSIP) (2015R1A2A1A15054248).}
\thanks{V.~V.~Mai was with Dankook University, Yongin 448-701, Republic of Korea. He is now with the Department of Automatic Control, KTH Royal Institute of Technology, SE-100 44 Stockholm, Sweden.
(E-mail: maivv@kth.se).}
\thanks{W.-Y. Shin (corresponding author) is with the Department of Computer Science and
Engineering, Dankook University, Yongin 448-701, Republic of Korea
(E-mail: wyshin@dankook.ac.kr).}
\thanks{K. Ishibashi is with the Advanced Wireless \& Communication Research Center (AWCC), The University of Electro-Communications, Tokyo 182-8585, Japan
(E-mail: koji@ieee.org).}
} \maketitle


\markboth{IEEE Journal of Selected Topics in Signal Processing}
{Mai \textit{\MakeLowercase{et al.}}: Wireless Power Transfer for
Distributed Estimation in Sensor Networks}


\begin{abstract}
This paper studies power allocation for distributed estimation of
an unknown scalar random source in  sensor networks with a
multiple-antenna fusion center (FC), where wireless sensors are
equipped with radio-frequency based energy harvesting technology.
The sensors' observation is locally processed by using an
\emph{uncoded amplify-and-forward} scheme.  The processed signals
are then sent to the FC, and are coherently combined at the FC, at
which the best linear unbiased estimator (BLUE) is adopted for
reliable estimation. We aim to solve the following two power
allocation problems: 1) minimizing distortion under various power
constraints; and 2) minimizing total transmit power under
distortion constraints, where the distortion is measured in terms
of mean-squared error of the BLUE. Two iterative algorithms are
developed to solve the non-convex problems, which converge at
least to a local optimum. In particular, the above algorithms are
designed to jointly optimize the \emph{amplification
coefficients}, \emph{energy beamforming}, and \emph{receive
filtering}. For each problem, a suboptimal design, a
single-antenna FC scenario, and a \emph{common harvester}
deployment for colocated sensors, are also studied. Using the
powerful \emph{semidefinite relaxation} framework, our result is
shown to be valid for any number of sensors, each with different
noise power, and for an arbitrarily number of antennas at the FC.
\end{abstract}


\newpage

\section{Introduction}
Distributed inference in wireless sensor networks (WSNs) has been
extensively studied for applications such as environmental
monitoring, weather forecasts, health care, and home automation
(see, e.g.,
\cite{YMG:08:CN,VV:97:PROC,XCL:08:SP,KV:13:IT,BEJ:14:SP,
JCS:14:SP,MJS:16:COML} and references therein).
Sensors in WSNs are  powered typically by batteries, and hence the
network lifetime is highly limited. In practice, periodically
replacing or recharging batteries may be hard or even impossible
(due to the fact that sensors are located inside toxic
environments,  building structures, or  human bodies
\cite{ZH:13:WCOM}). Therefore, although there have been many
efforts in power management policies, the network lifetime remains
a performance bottleneck and limits the wide-range deployment of
WSNs. 
\subsection{Previous Work}
The optimal power allocation strategies for distributed estimation
in WSNs have received a great research interest both from analog
and digital encoding perspectives
\cite{CXG:07:SP,ST:08:SP,XCL:08:SP,LA:09:SP,FL:09:SP,KV:13:IT,WW:13:WCOM,BEJ:14:SP}.
Among encoding schemes, the \emph{uncoded amplify-and-forward}
scheme has been extensively studied due to its simplicity and
information-theoretic-optimality properties under certain
conditions \cite{GAS:08:IT}.
In particular, the authors in \cite{CXG:07:SP} studied power
allocation for orthogonal multiple access channels (MACs), when
the best linear unbiased estimator (BLUE) is adopted. The same
problem was considered in \cite{XCL:08:SP} for a coherent MAC. The
effects of channel estimation error were reported in
\cite{ST:08:SP} for orthogonal MACs adopting a linear minimum
mean-squared error estimator, while in \cite{WW:13:WCOM}, the
sensing noise uncertainty was investigated by adopting the BLUE.
Recently, the optimal transmit strategy for cooperative linear
estimation  was studied in \cite{KV:13:IT}.

The tremendous performance gains achieved by multiple-antenna
techniques highly motivate us to integrate this technology into
future wireless systems including WSNs. The benefits of such
technology in the context of WSNs have been recently studied  for
distributed inference
\cite{BSTS:12:WCOM,BEJ:14:SP,NPC:14:SP,JCSL:15:SP,SDC:16:SP,MJS:16:COML}.
For a large-scale fusion center (FC) over a Rayleigh fading
channel, it has been shown in \cite{JCSL:15:SP}  that the
detection/estimation performance remains asymptotically constant
if the transmit power at each sensor decreases proportionally with
increasing number of antennas at the FC. The benefits of the
multiple-antenna FC in distributed detection were analyzed in
terms of asymptotic error exponents in \cite{MJS:16:COML}. Power
allocation strategies for distributed estimation  were studied for
the correlated source case \cite{SDC:16:SP} and  for the
correlated noise case \cite{BEJ:14:SP}.

Although the network life span  can be prolonged by applying the
aforementioned strategies, one needs a disruptive design that
fundamentally changes the limitation of a WSN. One of the
promising solution is  the so-called \emph{energy harvesting}
(EH), in which sensors scavenge energy from the ambient
environment (e.g., solar, wind, and vibration) that can  guarantee
an infinite life span in theory \cite{ NDA:15:JSAC}.
However, due to the unpredictable nature of energy sources,  EH is
typically uncontrolled, and thus  can be critical for some
reliable-sensitive applications. In addition to  commonly used
energy sources such as solar and wind, ambient radio-frequency
(RF) signals can be a viable new source for energy scavenging.
Most of the researches on \emph{wireless power transfer} (WPT)
have been focused on \emph{cellular} networks, where user
terminals replenish energy from the received signals sent by the
base station via the far-field RF-based WPT
\cite{GS:10:ISIT,ZH:13:WCOM, ZZH:13:COM,
SLZ:14:WCOM,XLZ:14:SP,LZC:14:SP,SLC:14:SP}. For example, the
fundamental trade-off between the achievable rate and the
transferred power was characterized
in \cite{GS:10:ISIT}. Several  practical receiver architectures
for simultaneous information and power transfer were investigated
in \cite{ZH:13:WCOM,ZZH:13:COM}. Exploiting multiple antenna
technologies in WPT has been widely studied:
multiple-input-multiple-output broadcast channels
\cite{ZH:13:WCOM},  beamforming designs for multiuser
multiple-input-single-output (MISO) \cite{XLZ:14:SP},
physical-layer security problems for multiuser MISO
\cite{LZC:14:SP}, and multiple-antenna interference channels
\cite{SLC:14:SP}. On the other hand, there are a relatively
limited number of studies on WPT for \emph{WSNs};  different WPT
technologies
for addressing energy/lifetime problems in WSNs were reviewed in
\cite{XSH:13:WC,KVB:14:PROC}; in \cite{YTG:15:GlobalSIP}, the
authors studied a distributed estimation system in which some of
the multiple-antenna sensors, named super sensors,  are capable of
WPT to its neighboring sensors via beamforming; and in
\cite{HH:15:GlobalSIP}, several multiple-antenna RF-based chargers
were used to replenish the wireless sensors and then to switch to
the information transmission phase, where each sensor sent a
quantized version of its measurement to the FC for estimation.
\subsection{Main Contributions}
For distributed estimation in WSNs, an important question is how
to intelligently exploit multiple-antenna technologies and WPT to
improve both the inference performance and network lifetime. In
this paper, we devote to studying the optimality of WPT and the
optimal  allocation of harvested energy for distributed estimation
of an unknown random source in WSNs with a multiple-antenna FC.
Our main contributions are  summarized as follows:
\begin{itemize}
    \item
 When the BLUE is adopted at the FC for estimation, we jointly optimize the amplification coefficients, energy beamforming, and receiver filtering by adopting alternative minimization methods (see Algorithms~\ref{alg:1} and \ref{alg:2}). To that end, we first solve the mean-squared error (MSE) minimization problem under the total power constraint at the FC as well as the causal power constraint at each sensor. Then, we solve a converse problem where the total transmit power at the FC is minimized subject to an MSE requirement.
    \item A key ingredient of our algorithms is the so-called \emph{semidefinite  relaxation}. We show that such a relaxation does not sacrifice the optimality of the relaxed problems. We derive the properties of the optimal solutions (see Theorems~\ref{thrm:1} and \ref{thrm:2}).
    \item A special deployment of WPT in WSNs is also discussed, where a common energy harvester is used to collect energy from the FC. We show that the optimization problems are significantly simplified in this case. The optimal power--distortion trade-off is also characterized (see Theorem~\ref{thrm:3}).
\end{itemize}
\subsection{Organization}
The rest of the paper is organized as follows. The system model
and problem formulation are described in
Section~\ref{sec:2}.Section~\ref{sec:3} studies the problem of
minimizing the MSE subject to power constraints. In
Section~\ref{sec:4}, the converse problem in Section~\ref{sec:3}
is studied. The numerical results are shown in
Section~\ref{sec:5}. Finally, we conclude the paper in
Section~\ref{sec:6}.
\subsection{Notations}
The operators $\left(\cdot\right)^\top$,
$\left(\cdot\right)^\ast$, $\left(\cdot\right)^\dag$ are the
transpose, complex conjugate, and transpose conjugate,
respectively. The notation $\IM{n}$ denotes the  $n \times n$
identity matrix; $\tr\left(\B{A}\right)$ denotes the trace of a
matrix $\B{A}$; $\rank{\B{A}}$ denotes the rank of a matrix
$\B{A}$; $\diag{{\B{a}}}$ denotes a diagonal matrix with vector
$\B{a}$ being its diagonal, $\B{A}\succeq\B{0}$ denotes the
positive semidefinite $\B{A}$; $\EX{\cdot}$ denotes the
expectation operator; $\textsf{dim}\left(\B{A}\right)$ denotes a
dimension of the subspace $\B{A}$. We use the Bachmann--Landau
notation:
 $        f\left(x\right)
                =O\left(g\left(x\right)\right)$ if $
                \lim_{x \to x_0}\frac{f\left(x\right)}{g\left(x\right)}=c < \infty$. Finally, we use the notation $\card{n}$ to denote the set of positive natural numbers up to $n$, i.e., $\card{n}=\{i:i=1,2,\ldots,n\}$.
\section{System Model and Problem Formulation}\label{sec:2}
\subsection{System Model}
\begin{figure}[t!]
\centerline{\includegraphics[width=0.82\textwidth]{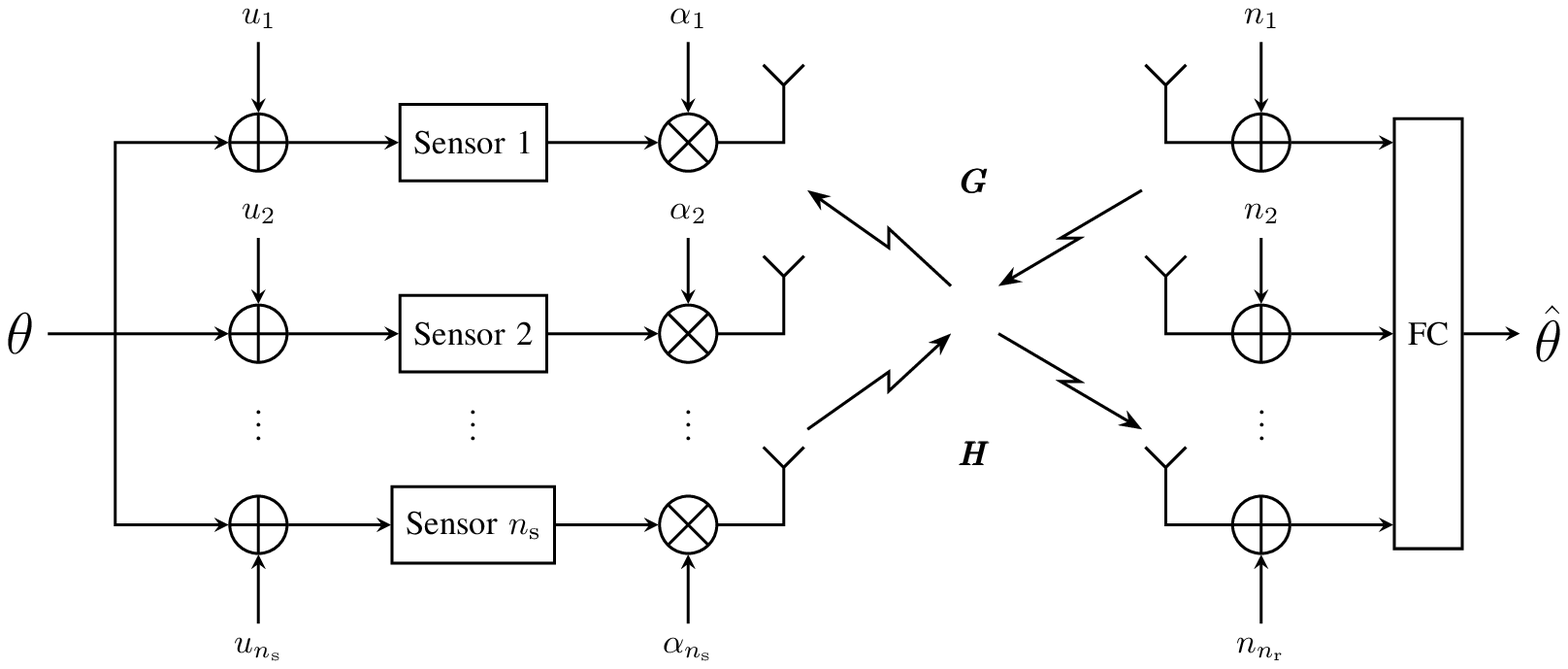}}
\caption{The distributed estimation system with an $\nRx$-antenna
FC and $\nSx$ spatially distributed sensors where
$\BB{G}=\left[\cvG{1}\,\,\cvG{2}\cdots\,\,\cvG{\nSx}\right]^\dag$.
} \label{fig:sm}
\end{figure}

As illustrated in Fig.~\ref{fig:sm}, we consider a distributed
estimation system where an $\nRx$-antenna FC collects data from
$\nSx$ spatially distributed sensors. Let $\theta$ be an unknown
scalar random parameter (source) with variance of $\Svar^2$ to be
estimated.\footnote{Our WPT framework is not designed only for a
scalar source. Note that there is no restriction to apply it for
the vector case even if finding a theoretical optimal solution
with no approximation for estimating vector-valued sources remains
an open problem.} Examples of such a parameter include physical
phenomena such as pressure, temperature, sound intensity,
radiation level, pollution concentration, seismic activity, etc..
We assume that all sensors do not have conventional energy
supplies and hence need to harvest energy from the RF signal
transferred by the FC for future use. We also assume that there is
no cooperation among the sensors since they are spatially
distributed. In this paper, we adopt a time-switching
\emph{harvest-then-forward} protocol \cite{LZC:14:COM} in which
for each $\tau\, T$ amount of time, where $T$ is the  length of
one time slot, the FC transmits its energy signal to the sensors,
and for the remaining $\left(1-\tau\right)T$ amount of time, the
sensors observe and forward their observations to the FC for
estimation while using the harvested energy from the RF signal.
For analytical convenience, we set $\tau=1/2$ in the sequel unless
otherwise specified.\footnote{For a Gaussian sensor network using
the source-channel encoding strategy\cite{GAS:08:IT}, the
rate-distortion theorems (e.g., \cite[Theorem 10.3.3]{CT:06:Book}
for orthogonal MACs and \cite[Section IV]{GAS:08:IT} for a
coherent MAC) enable us to characterize the effect of $\tau$ (via
the rate expressions) on the distortion performance (see, e.g.,
\cite{OGE:14:WCOM} and references therein). In this work, we adopt
an \emph{analog uncoded amplify-and-forward} scheme without
bandwidth expansion, in which the nature of information is in an
analog form, but not in a bitwise form \cite{GAS:08:IT}. As a
result, the power-distortion tradeoff (e.g.,
\cite[Theorem~1]{GV:03:Book}) is independent of $\tau$, and hence
in our work, the value of $\tau$ is assumed to be a constant. In
practice, the value $T_0=\left(1-\tau\right)T$ corresponds to the
amount of time that each sensor needs for observing, amplifying,
and forwarding its observation to the FC.}

In the first phase (i.e., the energy harvesting phase) of a time
slot, the FC broadcasts its energy signal to the sensors through
energy beamforming. More precisely, $\nBx \leq \nRx$ energy beams
are assigned to $\nSx$ sensors without loss of generality.  The
energy signal received at the $k$th sensor is then given  by
\begin{align}
    \RV{r}_k=\cvG{k}^\dag \BB{x}_{\mathrm{e}}+\RV{m}_k = \cvG{k}^\dag\sum_{i=1}^{\nBx}\tbf_i \RV{s}_i +\RV{m}_k,
\end{align}
where $\BB{x}_{\mathrm{e}}=\sum_{i=1}^{\nBx}\tbf_i \RV{s}_i$ is
the energy signal transmitted from the FC; $\RV{s}_i$ is the
energy-carrying signal for the $i$th energy beam fulfilling
$\mathds{E}\{\Norm{\RV{s}_i}^2\}=1$ and $\EX{\RV{s}_i \RV{s}_j}=0$
for $i \neq j$, which can be any arbitrary random signal provided
that its power spectral density satisfies certain regulations on
microwave radiation \cite{XLZ:14:SP}; $\tbf_i \in \C^{\nRx \times
1}$ is the $i$th energy bemforming vector; $\cvG{k} \in \C^{\nRx
\times 1}$ is the channel between the FC and $k$th sensor;  and
$\RV{m}_k$ is the additive noise at the $k$th sensor. By ignoring
the background noise for the sake of simplicity, the harvested
energy at the $k$th sensor in each slot is given by
\cite{ZH:13:WCOM}
 \begin{align}\label{eq:energy}
    E_k =  \frac{\zeta_k T}{2} \sum_{i=1}^{\nBx}\Norm{\tbf_i^\dag\cvG{k}}^2,
 \end{align}
 where $0\leq \zeta_k \leq 1$ is the energy harvesting efficiency at the $k$th sensor.
Then, the average power $P_k$ available for the information
transmission phase at the $k$th sensor can be expressed as
\begin{align}\label{eq:transmit:power}
    P_{k}=\frac{2\left(E_k-E^{\mathrm{cir}}_k\right)}{T}=\zeta_k \sum_{i=1}^{\nBx}\Norm{\tbf_i^\dag\cvG{k}}^2-\frac{2E^{\mathrm{cir}}_k}{T},
\end{align}
where $E_k^{\mathrm{cir}} \geq 0$ is the circuit energy
consumption at the $k$th sensor, which is assumed to be constant
over time slots. Similarly as in \cite{LZC:14:COM,SLC:14:SP}, we
simply assume $\zeta_k=1$ and unit slot duration in the rest of
this work (note that using an arbitrary $\zeta_k$ does not
fundamentally change our power allocation problems). Similarly as
in \cite{LZC:14:COM,JZ:14:WCOM}, for easy of presentation, we also
assume that $\{E_k^{\mathrm{cir}}\}_{k=1}^{\nSx}=0$ to focus on
the transmit power of the sensors.\footnote{Otherwise, we can
rewrite our problem along with a power offset, i.e.,
$E_k^{\mathrm{cir}}>0$, as a problem without any power offset for
a smaller $\zeta_k$.} The FC  has a total transmit power
constraint $P$; we thus have
 \begin{align}\label{eq:tot:const}
    \EX{\BB{x}_{\mathrm{e}}^\dag\BB{x}_{\mathrm{e}}}=\sum_{i=1}^{\nBx}\Vnorm{\tbf_i}^2 \leq P.
 \end{align}

Now, let us turn to describing the second phase (i.e., the
information transmission phase) of a time slot. The observation at
the $k$th sensor can be expressed as
\begin{align}
    x_k=\theta + u_k, \quad k=1,\ldots,\nSx,
\end{align}
where $u_k$ is the additive noise at the $k$th sensor with
variance $\Nvar{k}^2$. The noise at each sensor is assumed to be
independent of each other. In this paper, we adopt an analog
uncoded amplify-and-forward  scheme, i.e., the $k$th sensor just
simply amplifies its observation by a factor $\alpha_k$.
Therefore, by stacking the transmit signals from all sensors into
a single vector $\BB{t}$, it can be expressed as
\begin{align}\label{eq:transmit:signal}
    \BB{t}=\Ampl\B{1}\theta + \Ampl\uv,
\end{align}
where $\Ampl = \diag{{\ampl_1},{\ldots},{\ampl_{\nSx}}} \in
\C^{\nSx \times \nSx}$ is the amplification matrix; $\uv=\left[u_1
\, u_2 \, \cdots \, u_{\nSx} \right]^\top\in \C^{\nSx \times 1}$
is the noise vector at the sensors with zero mean and covariance
matrix
$\Rs=\diag{{\Nvar{1}^2},{\Nvar{2}^2},{\ldots},{\Nvar{\nSx}^2}}$;
and $\B{1}$ is the all one vector.  Then, the received signal $\zv
\in \C^{\nRx \times 1}$ at the FC can be written as
\begin{align}
    \zv=\ch\Ampl\B{1}\theta +  \ch\Ampl\uv +  \nv,
\end{align}
where $\ch \in \C^{\nRx \times \nSx}$ is the channel between the
sensors and FC;  and $\nv \in \C^{\nRx \times 1}$ is the noise
vector at the FC with zero mean and covariance matrix
$\Rn=\diag{{\RNvar{1}^2},{\RNvar{2}^2},{\ldots},{\RNvar{\nRx}^2}}$.
Here, the random quantities $\theta$, $\uv$, and $\nv$ are
statistically independent.

Since we consider a coherent MAC, we assume that there is perfect
synchronization between the sensors and the FC.
All wireless channels are assumed to be quasi-static flat fading,
i.e., once each channel is realized, it remains fixed during each
time slot and changes independently between slots. We further
assume that full channel state information (CSI) is available at
the FC. In practice,  the sensors-to-FC channel $\ch$ can be
estimated at the FC via periodic pilot transmissions from the
sensors, while the FC-to-sensors channels
$\{\cvG{i}\}_{i=1}^{\nSx}$ can be acquired owing to channel
reciprocity between the sensors-to-FC and FC-to-sensors channels
when the system operates in time-division-duplex mode.
\subsection{Problem Formulation}
The received signal $\zv$ is constructively combined at the FC by
a filtering vector $\rbf \in \C^{\nRx \times 1}$. Then, by
adopting the well-known BLUE \cite[Theorem~6.1]{Kay:93:Book}, the
FC estimates the parameter $\theta$ based on the minimal
sufficient statistic $\RV{y}=\rbf^\dag \zv$ as
follows:\footnote{The \emph{minimal sufficient statistic} is
defined in the sense that we no longer need the individual sample
since all the information has been captured by the sufficient
statistic \cite{Kay:93:Book}}.
 \begin{align}\label{eq:blue}
    \hat{\theta}
    =
        \left[
            \av^\dag
            \ch^\dag
            \rbf
            \RFCvar^{-2}
            \rbf^\dag
            \ch
            \av
        \right]^{-1}
        \av^\dag
        \ch^\dag
        \rbf
        \RFCvar^{-2}
        \RV{y},
\end{align}
where
$\RFCvar^{2}=\rbf^\dag\left[\ch\Ampl\Rs\Ampl^\dag\ch^\dag+\Rn\right]\rbf$
is the total noise power after post-processing at the FC; and
$\av=\left[\ampl_1\,\ampl_2\,\ldots \,\ampl_{\nSx}\right]^\top$.
The MSE of the BLUE can be written as
\begin{align}\label{eq:mse:def}
    \textsf{mse}
    &=
        \EX{|\theta-\hat{\theta}|^2}
    =
    \left[\av^\dag
            \ch^\dag
            \rbf
            \RFCvar^{-2}
            \rbf^\dag
            \ch
            \av
        \right]^{-1}
\nonumber\\
    &=
        \left[
            \frac{
                \Norm{\rbf^\dag\ch\av}^2
            }{
                \rbf^\dag
                \left(
                    \ch
                    \Ampl
                    \Rs
                    \Ampl^\dag
                    \ch^\dag
                    +
                    \Rn
                \right)
                \rbf
            }
        \right]^{-1}.
\end{align}

Since the three quantities $\av$, $\{\wv_k\}_{k=1}^{\nBx}$, and
$\rbf$ critically affect both the power requirement and estimation
performance of the entire system, we jointly design the optimum
sensor amplification coefficients $\av$, receive filtering vector
$\rbf$, and energy beamforming  $\{\wv_k\}_{k=1}^{\nBx}$ under
practical constraints. To that end, we solve two types of
minimization problems: 1) minimizing the MSE of the BLUE under
causal individual power constraints at the sensors and a total
power constraint at the FC; and 2) minimizing the total power
consumed at the FC given a minimum requirement of the MSE. In
particular, we aim to find the solution to the first problem,
named $(\textsf{P1})$, by solving the following optimization
problem.
\begin{align*}
\begin{aligned}
    &\Pb{1}:\\
    & \underset{\rbf,\av,\{\tbf_i\}_{i=1}^{\nBx}}{\text{maximize}}
    & & \frac{
                \Norm{\rbf^\dag\ch\av}^2
            }{
                \rbf^\dag
                \left(
                    \ch
                    \Ampl
                    \Rs
                    \Ampl^\dag
                    \ch^\dag
                    +
                    \Rn
                \right)
                \rbf
            }
 \\
    & \text{subject to}
    && \Norm{\alpha_k}^2\left(\Svar^2+\Nvar{k}^2\right) \leq  \sum_{i=1}^{\nBx}\Norm{\tbf_i^\dag\cvG{k}}^2, \quad \forall k \in \card{\nSx}
\\
    &&&\sum_{i=1}^{\nBx}\Vnorm{\tbf_i}^2 \leq P.
\end{aligned}
\end{align*}
As a counterpart of $\Pb{1}$, for a given MSE threshold
$\textsf{mse}=1/\gamma$, the second optimization problem is stated
as follows.
\begin{align*}
\begin{aligned}
    &\Pb{2}:\\
    & \underset{\rbf,\av,\{\tbf_i\}_{i=1}^{\nBx}}{\text{minimize}}
    & & \sum_{i=1}^{\nBx}\Vnorm{\tbf_i}^2
 \\
    & \text{subject to}
    && \frac{
                \Norm{\rbf^\dag\ch\av}^2
            }{
                \rbf^\dag
                \left(
                    \ch
                    \Ampl
                    \Rs
                    \Ampl^\dag
                    \ch^\dag
                    +
                    \Rn
                \right)
                \rbf
            } \geq \gamma,
\\
    &&& \Norm{\alpha_k}^2\left(\Svar^2+\Nvar{k}^2\right) \leq \sum_{i=1}^{\nBx}\Norm{\tbf_i^\dag\cvG{k}}^2, \quad \forall k \in \card{\nSx}.
\end{aligned}
\end{align*}


Note that closed-form solutions to the global optimization of
these two problems are generally unknown. Indeed, both  problems
are non-convex due to the coupled amplification vector $\av$ and
receive filtering $\rbf$. Therefore, we turn to a simple
approach---alternative minimization---which guarantees
convergence, at least to a local optimum.


\section{Minimizing MSE under Power Constraints}\label{sec:3}
In this section, we propose an alternative minimization algorithm
to obtain the minimum solution   to problem $\Pb{1}$. We also
study the MSE performance for a large-scale antenna FC as well as
a single-antenna FC.
\subsection{Proposed Solution to Problem $\Pb{1}$}
Since problem $\Pb{1}$ is non-convex due to a non-concave
objective function, we solve $\Pb{1}$ by using the alternative
minimization method. Our goal is to progressively increase the
objective function in $\Pb{1}$ by iteratively optimizing $\Pb{1}$
over $\av$ and $\{\tbf_i\}_{i=1}^{\nBx}$ for given $\rbf$, and
then over $\rbf$ for given $\av$. In order to find $\rbf$, we
first fix $\av$ and solve the following unconstrained optimization
problem:
\begin{align}\label{Pb:1:rbf}
\begin{aligned}
    & \underset{\rbf}{\text{maximize}}
    & & \frac{
                \Norm{\rbf^\dag\ch\av}^2
            }{
                \rbf^\dag
                \left(
                    \ch
                    \Ampl
                    \Rs
                    \Ampl^\dag
                    \ch^\dag
                    +
                    \Rn
                \right)
                \rbf
            },
\end{aligned}
\end{align}
which is a Rayleigh quotient and hence can be recasted as
\begin{align}\label{Pb:1:rbf}
\begin{aligned}
    & \underset{\rbf}{\text{minimize}}
    & & \rbf^\dag
                \left(
                    \ch
                    \Ampl
                    \Rs
                    \Ampl^\dag
                    \ch^\dag
                    +
                    \Rn
                \right)
                \rbf
 \\
    & \text{subject to}
    && \rbf^\dag\ch\av=1.
\end{aligned}
\end{align}
Solving the above problem, we obtain
\begin{align}\label{eq:rbf}
    \rbf^+=\kappa\left(
                    \ch
                    \Ampl
                    \Rs
                    \Ampl^\dag
                    \ch^\dag
                    +
                    \Rn
                \right)^{-1}
                \ch\av.
\end{align}
Note that the value of $\kappa$ is chosen to guarantee the
equality constraint in \eqref{Pb:1:rbf}. However, any selected
value of $\kappa$ will not affect the objective function in
$\Pb{1}$, and thus we simply choose $\kappa=1$ without loss of
optimality. For a given $\rbf$ in \eqref{eq:rbf}, we are now ready
to find an update of $\av$ and $\{\tbf_i\}_{i=1}^{\nBx}$ in
$\Pb{1}$.  To facilitate the calculations, we define
$\fv=\left[\rbf^\dag\cvH{1} \, \rbf^\dag\cvH{2} \, \ldots
\,\rbf^\dag\cvH{\nSx}\right]^\top$ and $\F=\diag{\fv}$, where
$\cvH{i}$ is the $i$th column of the matrix $\ch$. Then, for a
fixed receive filtering $\rbf$, problem $\Pb{1}$ can be expressed
as
\begin{align}\label{Pb:1:1}
\begin{aligned}
    &\hspace{-0.2cm} \underset{\av,\{\tbf_i\}_{i=1}^{\nBx}}{\text{maximize}} \hspace{-0.1cm}
    & & \frac{
                \Norm{\av^\top\fv}^2
            }{
                \av^\top
                \F
                \Rs
                \F^\dag
                \av^*
                +
            \rbf^\dag\Rn\rbf
            }
 \\
    &\hspace{-0.2cm} \text{subject to} \hspace{-0.1cm}
    && \Norm{\alpha_k}^2\left(\Svar^2+\Nvar{k}^2\right) \leq \sum_{i=1}^{\nBx}\Norm{\tbf_i^\dag\cvG{k}}^2, \, \forall k \in \card{\nSx}
\\
    &&&\sum_{i=1}^{\nBx}\Vnorm{\tbf_i}^2 \leq P.
\end{aligned}
\end{align}
We remark that even with a fixed receive filtering $\rbf$, problem
$\Pb{1}$ is still non-convex, and thus needs to be transformed to
a simple form. We further introduce $\Q=\av^* \av^\top$,
$\W=\sum_{i=1}^{\nBx}\tbf_i\tbf_i^\dag$, $\B{\Sigma}=\fv\fv^\dag$,
$\B{\Psi}=\F\Rs\F^\dag$, $\G_k=\cvG{k}\cvG{k}^\dag$, and
$\D_k=\mathrm{diag}({0},\ldots,{\Svar^2+\Nvar{k}^2},\ldots,0)$.
Then, we can rewrite the optimization problem \eqref{Pb:1:1} as
\begin{align}\label{Pb:1:2}
\begin{aligned}
    & \underset{\Q,\W}{\text{maximize}}
    & & \frac{
                \tr\left(\Q\B{\Sigma}\right)
            }{
                \tr\left(
                    \Q\B{\Psi}
                \right)
                    +
                    \rbf^\dag\Rn\rbf
            }
 \\
    & \text{subject to}
    && \tr\left(\D_k\Q\right)-\tr\left(\G_k\W\right) \leq 0, \quad \forall k \in \card{\nSx}
\\
    &&& \tr\left(\W\right) \leq P
\\
    &&& \W \succeq \B{0},  \Q \succeq \B{0}
\\
    &&& \rank{\Q}=1.
\end{aligned}
\end{align}
In \eqref{Pb:1:2}, if there exist a rank one solution of the
optimal $\Q=\Q^\star$ and a rank $\nBx$ solution of the optimal
$\W=\W^\star$, then one can recover the optimal $\av^\star$ and
$\{\tbf_i^\star\}_{i=1}^{\nBx}$ by taking the eigenvalue
decomposition of the matrices $\Q^\star$ and $\W^\star$,
respectively. Note that  problem \eqref{Pb:1:2} is non-convex due
to  the linear fractional structure of its objective function.
However, we can use the Charnes-Cooper transformation
\cite{CC:62:NRLQ} to reformulate the quasi-convex objective
function in \eqref{Pb:1:2} into a simpler form as
follows:\footnote{Here, we use the transformations
$\eta^{-1}=\tr\left(\left(\Q\B{\Psi}\right)+\rbf^\dag\Rn\rbf\right)$,
$\nQ=\eta\Q$, and $\nW=\eta\W$.}
\begin{align}\label{Pb:1:3}
\begin{aligned}
    & \underset{\nQ,\nW,\eta}{\text{maximize}}
    &&  \tr\left(\nQ\B{\Sigma}\right)
\\
    & \text{subject to}
    &&\tr\left(\nQ\B{\Psi}\right)+\eta\rbf^\dag\Rn\rbf=1
\\
    &&& \tr\left(\D_k\nQ\right)-\tr\left(\G_k\nW\right) \leq 0, \quad \forall k \in \card{\nSx}
\\
    &&& \tr\left(\nW\right) \leq \eta P
\\
    &&& \nW \succeq \B{0},  \nQ \succeq \B{0}, \eta > 0
\\
    &&& \rank{\nQ}=1.
\end{aligned}
\end{align}
Note that $\eta=0$ is not feasible because from the third
constraint, we must have $\nW=\BB{0}$ if $\eta=0$. Thus, from the
second constraint for any $k$, it follows that $\nQ=\BB{0}$, which
however violates the first constraint in \eqref{Pb:1:3}.
\begin{remark}[The Equivalence of Problems \eqref{Pb:1:2} and \eqref{Pb:1:3}]
If $(\nQ^\star,\nW^\star,\eta^\star)$ is the optimal solution to
problem \eqref{Pb:1:3}, then
$(\nQ^\star/\eta^\star,\nW^\star/\eta^\star)$ is feasible to
problem \eqref{Pb:1:2} and achieves the same objective value as
that of problem \eqref{Pb:1:3}. On the other hand, let
$t^\star=\tr\left(\Q^\star\B{\Psi}\right)+\rbf^\dag\Rn\rbf$. Then,
if $(\Q^\star,\W^\star)$ is the optimal solution to  problem
\eqref{Pb:1:2}, then
$(\Q^\star/t^\star,\W^\star/t^\star,1/t^\star)$ is feasible to
problem \eqref{Pb:1:3} and achieves the same objective value as
that of  problem \eqref{Pb:1:2}. This implies that the
Charnes-Cooper transform is a one-to-one mapping between the
feasible sets of problems \eqref{Pb:1:2} and \eqref{Pb:1:3}. We
can thus obtain the optimal solution to problem \eqref{Pb:1:2} by
solving  problem \eqref{Pb:1:3}, which has a simpler form in the
sense that the non-convexity of the objective function in problem
\eqref{Pb:1:2} is eliminated.
\end{remark}

Note that problem \eqref{Pb:1:3} is still non-convex due to the
rank constraint, which makes  problem \eqref{Pb:1:3} intractable
in general. Hence, we will solve a relaxed version of
\eqref{Pb:1:3} by ignoring the rank constraint on $\Q$, which
leads to the semidefinite relaxation (SDR) of  problem
\eqref{Pb:1:3}.
\begin{align*}
\begin{aligned}
    &\sdr{1}:\\
    & \underset{\nQ,\nW,\eta}{\text{maximize}}
    &&  \tr\left(\nQ\B{\Sigma}\right)
\\
    & \text{subject to}
    &&\tr\left(\nQ\B{\Psi}\right)+\eta\rbf^\dag\Rn\rbf=1
\\
    &&& \tr\left(\D_k\nQ\right)-\tr\left(\G_k\nW\right) \leq 0, \,\,\, \forall k \in \card{\nSx}
\\
    &&& \tr\left(\nW\right) \leq \eta P
\\
    &&& \nW \succeq \B{0},  \nQ \succeq \B{0}, \eta > 0.
\end{aligned}
\end{align*}
The relaxed problem $\sdr{1}$ is now convex---indeed semidefinite
program (SDP)---whose optimal solution can be found, for example,
by using the interior-point method (e.g., CVX \cite{cvx}). The
following theorem characterizes the properties of the optimal
solution to problem~$\sdr{1}$.

\begin{theorem}[Properties of Optimal Solution]\label{thrm:1}
Let $\nu^\star$ and $\beta^\star$ be the optimal dual solutions
associated with the first and third constraint in $\sdr{1}$,
respectively. We also let $\nQ^\star$ and $\nW^\star$ be the
optimal primal solutions to problem $\sdr{1}$. Then, the following
three properties are fulfilled:
\begin{enumerate}
\item $\nu^\star >0$, $\beta^\star>0$; \item
$\rank{\nW^\star}\leq\min\left(\nSx,\nRx\right)$; \item
$\rank{\nQ^\star}=1$.
\end{enumerate}
\begin{proof}
See Appendix~\ref{appdx:1}.
\end{proof}
\end{theorem}

\begin{remark}
The condition $\beta^\star>0$ implies that the total power
constraint at the FC must be satisfied with equality, while
property 2) implies that at most $\nBx=\min\left(\nSx,\nRx\right)$
energy beams are required for the optimal solution of problem
$\sdr{1}$. It is worth noting that for fixed $\rbf$, at the
optimal solution $(\nQ^\star,\nW^\star,\eta^\star)$, the
individual power constraints in $\sdr{1}$ are not necessarily all
tight, i.e., there may exist some $k$ such that
$\tr\left(\D_k\nQ\right)-\tr\left(\G_k\nW\right) < 0$. This fact
reveals that the sensors do not always transmit all the power
budget harvested from the energy harvesting phase, but power
control is required to guarantee the MSE optimality. A similar
observation was made in throughput optimization for
multiple-antenna multiuser cellular systems in \cite{LZC:14:COM}.
\end{remark}

\begin{remark}[The Equivalence of Problems \eqref{Pb:1:3} and $\sdr{1}$]\label{rmk:eqv:1:1:3:sdr}
We remark that since problem $\sdr{1}$ is a relaxed version of
problem \eqref{Pb:1:3}, in general, the solution to problem
$\sdr{1}$ provides an upper bound on the optimal solution to
problem \eqref{Pb:1:3}, or equivalently, an upper bound on problem
$\Pb{1}$ for a given $\rbf$. Fortunately, we can show that the
optimal solution to $\sdr{1}$ is also optimal to \eqref{Pb:1:3}.
To do that, let $\Pf{\eta}{\nQ,\nW}$ be the objective function of
problem  \eqref{Pb:1:3} or $\sdr{1}$ for a given feasible $\eta$,
and $(\nQ^\star,\nW^\star)$ and $(\nQ_\star,\nW_\star)$ be the
optimal solutions to problems $\sdr{1}$ and \eqref{Pb:1:3},
respectively. Since the optimization problem $\sdr{1}$ is a
relaxation of problem \eqref{Pb:1:3}, we must have
\begin{align}\label{eq:rmk:1}
    \Pf{\eta}{\nQ^\star,\nW^\star} \geq \Pf{\eta}{\nQ_\star,\nW_\star}.
\end{align}
On the other hand, since $\rank{\nQ^\star}=1$, the solution
$(\nQ^\star,\nW^\star,\eta)$ is also a feasible solution to
problem \eqref{Pb:1:3}. Therefore, we have
\begin{align}\label{eq:rmk:2}
    \Pf{\eta}{\nQ^\star,\nW^\star} \leq \Pf{\eta}{\nQ_\star,\nW_\star}.
\end{align}
From \eqref{eq:rmk:1} and \eqref{eq:rmk:2}, it follows that
$\Pf{\eta}{\nQ^\star,\nW^\star} = \Pf{\eta}{\nQ_\star,\nW_\star}$.
In other words, $(\nQ^\star,\nW^\star)$ is also optimal solution
to problem \eqref{Pb:1:3}. Note that the above equivalence holds
for any feasible $\eta$, and hence it holds for the optimal
$\eta$.
\end{remark}

 Remark~\ref{rmk:eqv:1:1:3:sdr} suggests that we can solve the original problem $\Pb{1}$ for a given $\rbf$ by equivalently solving the relaxed problem $\sdr{1}$ without loss of optimality. Finally, we  summarize the overall procedure for solving problem $\Pb{1}$ in Algorithm~\ref{alg:1} below. In this algorithm, the FC iteratively updates $\rbf$, $\av$ and $\{\tbf_i\}_{i=1}^{\nBx}$ in Step~3 and 4, respectively. The convergence and complexity of Algorithm~\ref{alg:1} are analyzed in the following remark.

\begin{remark}[Convergence and Complexity]\label{rmk:convergence} Note that the objective function in $\Pb{1}$ is increased in each step of Algorithm~\ref{alg:1}. Moreover, the objective function is upper-bounded by a certain value due to the finite total power at the FC, which implies that the algorithm must converge. However, the algorithm may converge to a local optimum due to the non-convex nature of the optimization problem. We now provide the complexity analysis of the proposed algorithm. Specifically, in each iteration of Algorithm~\ref{alg:1}, the worst-case computational complexity for solving the generic convex problem in $\sdr{1}$, corresponding to Step 4 in Algorithm~\ref{alg:1}, using the interior point method is given by $O\left(\left(2 \nSx+\nRx\right)^{1/2}\left(\nSx^4 + \nRx^3\nSx + \nRx^2\nSx^2\right)\log\left(\frac{1}{\xi}\right)\right)$ for an $\xi$-optimal solution \cite[Chapter 6.6.3]{AN:01:Book}. For the receive filtering update, based on the elementary vector matrix calculation \cite{Hun:07:tchrp}, one can show that the computational complexity of Step 3 in Algorithm~\ref{alg:1} is $O\left(\nRx^3+\nRx^2 \nSx\right)$.
In Step 6, the amplification vector is constructed by using
eigenvalue decomposition of rank one matrix $\nQ$, and hence the
complexity is $O\left(\nSx^2\right)$. Thus, the overall complexity
per iteration of Algorithm~\ref{alg:1} is at most $O\left(\left(2
\nSx+\nRx\right)^{1/2}\left(\nSx^4 + \nRx^3\nSx +
\nRx^2\nSx^2\right)\log\left(\frac{1}{\xi}\right)+\nRx^3+\nRx^2
\nSx\right)$. We remark that although the complexity of the
alternative minimization algorithms are typically unknown
\cite{GCJ:11:IT,SRL:11:SP}, it is observed via simulations that
they converge within 10 to 20 iterations in general.
\end{remark}

\subsection{Large-Scale Antenna FC}
The following proposition shows the property of the asymptotic MSE
of the BLUE.
\begin{proposition}[Asymptotic MSE]\label{prop:asym}
Consider the distributed estimation system in Section~\ref{sec:2},
where the channel matrix $\ch$ is a random matrix with independent
and identical elements, each of which has zero mean and unit
variance. As the number of antennas at the FC tends to infinity,
the MSE defined in \eqref{eq:mse:def} converges to that of
centralized estimation systems.\footnote{We use the term
centralized estimation to refer to the case for which the sensors'
data are perfectly available at the FC, which serves as a
performance benchmark.} That is, as $\nRx \to \infty$, we have
\begin{align}\label{eq:mse:bmk}
    \textsf{mse}
     \overset{\textsf{a.s}}\to
    \left[\B{1}^\top \Rs^{-1} \B{1}\right]^{-1},
\end{align}
where $\overset{\textsf{a.s}} \to$ denotes the almost sure
convergence.
\begin{proof}
Given the receive filtering in \eqref{eq:rbf}, the MSE can be
written as
\begin{align}\label{eq:prop1:eq1}
    \textsf{mse}
    = \left[
        \av^\dag
        \ch^\dag
        \left(
            \ch
            \Ampl
            \Rs
            \Ampl^\dag
            \ch^\dag
            +
            \Rn
        \right)^{-1}
        \ch\av
    \right]^{-1}.
\end{align}
Therefore, it suffices to prove that
\begin{align}
        \av^\dag
        \ch^\dag
        \left(
            \ch
            \Ampl
            \Rs
            \Ampl^\dag
            \ch^\dag
            +
            \Rn
        \right)^{-1}
        \ch\av
         \overset{\textsf{a.s}} \to
        \B{1}^\top \Rs^{-1} \B{1}
\end{align}
as $\nRx$ tends to infinity. Using the matrix inversion lemma, we
can show that
\begin{align}\label{eq:prop1:eq2}
    &\left(
            \ch
            \Ampl
            \Rs
            \Ampl^\dag
            \ch^\dag
            +
            \Rn
        \right)^{-1}
\nonumber\\
&\hspace{0.25cm}
    =
    \Rn^{-1}
    -
    \Rn^{-1}
    \ch
    \left(
        \BB{K}^{-1}
        +
        \ch^\dag\Rn^{-1}
        \ch
    \right)^{-1}
    \ch^\dag
    \Rn^{-1},
\end{align}
where $\BB{K}=\Ampl\Rs\Ampl^\dag$. Substituting
\eqref{eq:prop1:eq2} into \eqref{eq:prop1:eq1}, we obtain
\begin{align}
    \textsf{mse}^{-1}
    &=
    \av^\dag
    \ch^\dag
    \Rn^{-1}
    \ch
    \av
    \nonumber\\
    &\hspace{-0.5cm}
    -
    \av^\dag
    \ch^\dag
    \Rn^{-1}
    \ch
    \left(
        \BB{K}^{-1}
        +
        \ch^\dag\Rn^{-1}
        \ch
    \right)^{-1}
    \ch^\dag
    \Rn^{-1}
    \ch
    \av.
\end{align}
Note that as $\nRx \to \infty$, we have \cite{Mar:10:WCOM}
\begin{align}
    \frac{1}{\nRx}\ch^\dag\Rn^{-1}\ch \overset{\textsf{a.s}} \to \Rn^{-1}.
\end{align}
Using this identity, we obtain
\begin{align}\label{eq:asym:mse}
    \frac{\textsf{mse}^{-1}}{\nRx}
    & \overset{\textsf{a.s}} \to
        \av^\dag
        \Rn^{-1}
        \av
        -
        \av^\dag
        \Rn^{-1}
        \left(
            \frac{\BB{K}^{-1}}{\nRx}
            +
            \Rn^{-1}
        \right)^{-1}
        \Rn^{-1}
        \av
    \nonumber\\
    &=
        \av^\dag
        \Rn^{-1}
        \left(
            \IM{}
            -
            \left(
                \frac{\Rn\BB{K}^{-1}}{\nRx}
                +
                \IM{}
            \right)^{-1}
        \right)
        \av
    \nonumber\\
    &=
        \frac{1}{\nRx}
        \av^\dag
        \left(\BB{K}+\frac{\Rn}{\nRx}\right)^{-1}
        \av.
\end{align}
where the second equality follows from the matrix inversion lemma.
From \eqref{eq:asym:mse} and the definitions of the matrices
$\BB{K}$ and $\Ampl$, as $\nRx \to \infty$, we finally have
\begin{align}
    \textsf{mse}
     \overset{\textsf{a.s}}\to
    \left[\B{1}^\top \Rs^{-1} \B{1}\right]^{-1},
\end{align}
which concludes the proof of the proposition.
\end{proof}
\end{proposition}
\begin{algorithm}[t]
\caption{ proposed algorithm to solve $\Pb{1}$}
 \begin{algorithmic}[1]\label{alg:1}
\STATE \textbf{Initialization}: set $n :=0$, and generate $\av^{(0)}$ and  $\Ampl^{(0)}$.\\[0.15cm]

\STATE \textbf{repeat}\\[0.12cm]

\STATE \quad $\rbf^{(n)}=\left(\ch\Ampl^{(n)}\Rs{\Ampl^{(n)}}^\dag\ch^\dag+\Rn\right)^{-1}\ch\av^{(n)}$\\[0.12cm]

\STATE \quad Solve problem $\sdr{1}$ with $\rbf=\rbf^{(n)}$ to obtain the \\
\quad  optimal solution $(\nQ^\star,\nW^\star,\eta^\star)$.\\[0.12cm]

\STATE \quad Set $(\nQ^{(n+1)},\nW^{(n+1)},\eta^{(n+1)}):=(\nQ^\star,\nW^\star,\eta^\star)$.\\[0.12cm]

\STATE \quad Construct $\{\av^{(n+1)},\Ampl^{(n+1)}\}$ from $\nQ^{(n+1)}/\eta^{(n+1)}$.\\[0.12cm]

\STATE \quad Update $n := n+1$.\\[0.12cm]

\STATE \textbf{until convergence}\\[0.12cm]
\STATE \textbf{Output:}
$(\Q=\nQ^{(n)}/\eta^{(n)},\W=\nW^{(n)}/\eta^{(n)},\rbf=\rbf^{(n)})$
\end{algorithmic}
\end{algorithm}

Proposition~\ref{prop:asym} implies that as the number of antennas
grows large, the effects of fading and noise at the FC disappear,
and hence the performance benchmark is determined by the sensing
quality. From \eqref{eq:mse:bmk}, if the sensing noise at the
sensors is equal to $\Rs=\sigma_{\mathrm{n}}^2\IM{\nSx}$, then it
follows that $\left[\B{1}^\top \Rs^{-1}
\B{1}\right]^{-1}=\frac{\sigma_{\mathrm{n}}^2}{\nSx}$. This means
that the MSE linearly decreases according to $\nSx$.

\subsection{Single-Antenna FC}
It is of importance to study the single-antenna FC scenario
separately not only for comparison but also because the problem is
remarkably simplified. Specifically, for a single-antenna FC, the
design of energy beamforming and receive filtering is neglected,
and thus we aim to simply find the optimal amplification
coefficients $\av$ that minimize the MSE of the BLUE. During the
energy transmission phase, we assume that the FC transmits an
energy signal $\RV{s}$ such that
$\mathds{E}\{\Norm{\RV{s}}^2\}=P$. In this case, the harvested
energy at the $k$th sensor is given by
\begin{align}
    E_k=\frac{P\Norm{\RV{g}_k}^2}{2}.
\end{align}
The MSE of the BLUE is boiled down to
\begin{align}\label{eq:mse:def:fc}
    \textsf{mse}
    =
        \left[
            \frac{
                \av^\top\cvH{}\cvH{}^\dag\av^*
            }{
                \av^\top
                \F
                \Rs
                \F^\dag
                \av^*
                +
                \nvar^2
            }
        \right]^{-1},
\end{align}
where $\cvH{}^\dag \in \C^{1 \times \nSx}$ is the channel between
the sensors and the FC; $\F=\diag{{\cvH{}}}$; and $n \in \C$ is
the additive noise at the FC. Given the MSE in
\eqref{eq:mse:def:fc}, we aim to solve the following problem
\begin{align}\label{Pb:mse:fc}
\begin{aligned}
    & \underset{\av}{\text{maximize}}
    & & \frac{
                \av^\top\cvH{}\cvH{}^\dag\av^*
            }{
                \av^\top
                \F
                \Rs
                \F^\dag
                \av^*
                +
                \nvar^2
            }
 \\
    & \text{subject to}
    && \Norm{\alpha_k}^2\left(\Svar^2+\Nvar{k}^2\right) \leq P\Norm{\RV{g}_k}^2, \,\,\,\, \forall k \in \card{\nSx}.
\end{aligned}
\end{align}
The above problem---quadratically constrained ratio of quadratic
functions (QCRQ)---has been studied for parameter tracking using
the Kalman filter at the FC \cite{JCS:14:SP}, where the optimal
solution is given by
\begin{align}\label{eq:av:fc}
    \av^\star=\frac{1}{\sqrt{\mE{\BB{P}^\star}{\nSx+1}{\nSx+1}}}\bar{\av}^*.
\end{align}
Here, $\mE{\BB{P}^\star}{i}{j}$ is the $\left(i,j\right)$th
element of the matrix $\BB{P}^\star$; $\bar{\av}$ is the vector
satisfying $\bar{\av}^*\bar{\av}^\top=\BB{P}_{\nSx}^\star$;
$\BB{P}_{\nSx}^\star$  is the $\nSx$th order leading principal
submatrix of $\BB{P}^\star$ obtained by excluding the $(\nSx+1)$th
row and column; and $\BB{P}^\star  \in
\C^{\left(\nSx+1\right)\times \left(\nSx+1\right)}$ is the optimal
solution to the following problem
 \begin{align}\label{Pb:mse:fc:sdp}
\begin{aligned}
    & \underset{\BB{P} \succeq 0}{\text{maximize}}
    & & \tr\left(\BB{P}\,\BB{\Xi}\right)
 \\
    & \text{subject to}
    && \tr\left(\BB{P}\,\BB{C}\right)=1 \\
    &&& \tr\left(\BB{P}\,\bar{\D}_k\right) \leq P\Norm{\RV{g}_k}^2, \quad \forall k \in \card{\nSx},
\end{aligned}
\end{align}
where $\BB{P}=\left[t\,\av^\top\,\,
t\right]^\dag\left[t\,\av^\top\,\, t\right]$; $ \BB{\Xi}=\left(
  \begin{array}{cc}
   \cvH{}\cvH{}^\dag  & \B{0} \\
    \B{0} & 0 \\
  \end{array}
\right) $; $\BB{C}=\left(
  \begin{array}{cc}
    \F\Rs\F^\dag & \B{0} \\
    \B{0} & t\,\nvar^2 \\
  \end{array}
\right) $; $ \bar{\D}_k=\left(
  \begin{array}{cc}
    \D_k & \B{0} \\
    \B{0} & -P\Norm{\RV{g}_k}^2 \\
  \end{array}
\right)$; and $t$ is an auxiliary variable.

\begin{remark}
Note that the solution in \eqref{eq:av:fc} is indeed global
optimum. This is different from the multiple-antenna FC case in
which we may only achieve a local optimum.
\end{remark}
\subsection{A Common Energy Harvester}
We now consider a special deployment case in WPT-enabled sensor
networks, where a common energy harvester is used to collect
energy from the FC.\footnote{This differs from what we have
considered so far, where each sensor has its own energy harvester.
This approach reduces the hardware complexity of  sensors.
However, it is feasible only if the sensors are closely, or even
colocated.} Assume that the common energy harvester is equipped
with a single antenna,  then the optimization problem can be
stated as
\begin{align*}
\begin{aligned}
     \PbS{1}: \quad & \underset{\rbf,\av,\tbf}{\text{maximize}}
    & & \frac{
                \Norm{\rbf^\dag\ch\av}^2
            }{
                \rbf^\dag
                \left(
                    \ch
                    \Ampl
                    \Rs
                    \Ampl^\dag
                    \ch^\dag
                    +
                    \Rn
                \right)
                \rbf
            }
 \\
    & \text{subject to}
    && \av^\dag\D \av \leq \Norm{\tbf^\dag \he}^2
\\
    &&&\Vnorm{\tbf}^2 \leq P,
\end{aligned}
\end{align*}
where $\he$ is the channel between the FC and the common harvester
and
$\D=\diag{{\Svar^2+\Nvar{1}^2},{\cdots},{\Svar^2+\Nvar{\nSx}^2}}$.
Since the objective function in problem $\PbS{1}$ is monotonically
increasing with the norm of $\av$, the sum power constraint (i.e.,
the first constraint) should be satisfied with equality and the
right-hand side of this constraint should be as large as possible
to be the optimal solution. This implies that the optimal energy
beamforming is $\wv^\star=\sqrt{P}\frac{\he}{\Vnorm{\he}}$.
Similarly as in problem $\Pb{1}$, we iteratively solve the above
problem for $\rbf$ and $\av$, where $\rbf$ is given in
\eqref{eq:rbf}. Let $E=\frac{1}{2}|\he^\dag
\wv^\star|^2=\frac{1}{2}P\Vnorm{\he}^2$ be the harvested energy at
the harvester. For a given $\rbf$, the optimal $\av$ is then the
solution of the following problem:
\begin{align}\label{Pb:mse:sum:av}
\begin{aligned}
    & \underset{\av}{\text{maximize}}
    & & \frac{
                \Norm{\av^\top\fv}^2
            }{
                \av^\top
                \F
                \Rs
                \F^\dag
                \av^*
                +
            \rbf^\dag\Rn\rbf
            }
 \\
    & \text{subject to}
    && \av^\dag\D \av = P\Vnorm{\he}^2,
\end{aligned}
\end{align}
where $\fv=\left[\rbf^\dag\cvH{1} \, \rbf^\dag\cvH{2} \, \cdots
\,\rbf^\dag\cvH{\nSx}\right]^\top$ and $\F=\diag{\fv}$. The
problem is equivalent to
\begin{align}\label{Pb:mse:sum:av:Rayleigh}
\begin{aligned}
    & \underset{\av}{\text{maximize}}
    & & \frac{
                \av^\top \BB{X} \av^*
            }{
                \av^\top
                \BB{Y}
                \av^*
            },
\end{aligned}
\end{align}
where $\BB{X}=\fv\fv^\dag$ and
$\BB{Y}=\F\Rs\F^\dag+\frac{\rbf^\dag\Rn\rbf}{P\Vnorm{\he}^2}\D$.
Note that $\BB{X} \succeq \B{0}$ and $\BB{Y} \succ \B{0}$, and
problem \eqref{Pb:mse:sum:av:Rayleigh} is indeed Rayleigh
quotient, thus the optimal solution can be expressed as
\begin{align}
    \av^\star
    =
    \sqrt{\frac{P\Vnorm{\he}^2}{\fv^\dag\BB{Y}^{-1}\D\BB{Y}^{-1}\fv}}
    \BB{Y}^{-1}
    \fv^*.
\end{align}
Then, the optimal value in \eqref{Pb:mse:sum:av} is given by
\begin{align}\label{eq:mse:sum:obj}
    \hspace{-0.3cm}
    \max_{\av} \frac{
                \Norm{\av^\top\fv}^2
            }{
                \av^\top
                \F
                \Rs
                \F^\dag
                \av^*
                +
            \rbf^\dag\Rn\rbf
            }
    =
        \maxeig\left(\BB{Y}^{-1}\BB{X}\right)
    =    \fv^\dag \BB{Y}^{-1} \fv,
\end{align}
where $\maxeig\left(\cdot\right)$ denotes the maximum eigenvalue
of a matrix. It can be seen that the sum power constraint enables
to significantly reduce the complexity of the optimization
problem.
\section{Minimizing Power under an MSE Constraint}\label{sec:4}
In this section, we study the power minimization for distributed
estimation with an MSE constraint. 
\subsection{Proposed Solution to Problem $\Pb{2}$}
Similarly as in problem $\Pb{1}$, we adopt an alternative
minimization method to iteratively solve problem $\Pb{2}$.
Specifically, we first solve problem $\Pb{2}$ over $\rbf$ for
given $\av$ by finding a solution to the following feasibility
problem:
\begin{align}\label{Pb:2:rbf}
\begin{aligned}
    & \underset{\rbf}{\text{minimize}}
    & & 0
 \\
    & \text{subject to}
    && \frac{
                \Norm{\rbf^\dag\ch\av}^2
            }{
                \rbf^\dag
                \left(
                    \ch
                    \Ampl
                    \Rs
                    \Ampl^\dag
                    \ch^\dag
                    +
                    \Rn
                \right)
                \rbf
            } \geq \gamma.
\end{aligned}
\end{align}
Since the left-hand side (LHS) of the constraint in
\eqref{Pb:2:rbf} is increasing with the norm of $\av$, one should
choose $\rbf$ such that the LHS term is as large as possible.
Hence, problem \eqref{Pb:2:rbf} can be  rewritten as an
unconstrained optimization problem as follows:
\begin{align}\label{Pb:2:rbf:non-cst}
\begin{aligned}
    & \underset{\rbf}{\text{maximize}}
    & & \frac{
                \Norm{\rbf^\dag\ch\av}^2
            }{
                \rbf^\dag
                \left(
                    \ch
                    \Ampl
                    \Rs
                    \Ampl^\dag
                    \ch^\dag
                    +
                    \Rn
                \right)
                \rbf
            }.
\end{aligned}
\end{align}
Solving the above problem, we obtain
\begin{align}\label{eq:rbf:P2}
    \rbf^+=\left(
                    \ch
                    \Ampl
                    \Rs
                    \Ampl^\dag
                    \ch^\dag
                    +
                    \Rn
                \right)^{-1}
                \ch\av.
\end{align}
For fixed $\rbf$ given in \eqref{eq:rbf:P2}, we now solve problem
$\Pb{2}$ over $\av$ and $\{\tbf_i\}_{i=1}^{\nBx}$ as in the
following:
\begin{align}\label{Pb:2:1}
\begin{aligned}
    &\hspace{-0.2cm} \underset{\av,\{\tbf_i\}_{i=1}^{\nBx}}{\text{minimize}} \hspace{-0.1cm}
    & & \sum_{i=1}^{\nBx}\Vnorm{\tbf_i}^2
 \\
    & \hspace{-0.2cm} \text{subject to}  \hspace{-0.1cm}
    && \frac{
                \Norm{\av^\top\fv}^2
            }{
                \av^\top
                \F
                \Rs
                \F^\dag
                \av^*
                +
            \rbf^\dag\Rn\rbf
            }            \geq \gamma,
\\
    &&& \hspace{-0.12cm} \Norm{\alpha_k}^2\left(\Svar^2+\Nvar{k}^2\right) \leq \sum_{i=1}^{\nBx}\Norm{\tbf_i^\dag\cvG{k}}^2, \, \forall k \in \card{\nSx},
\end{aligned}
\end{align}
where $\fv=\left[\rbf^\dag\cvH{1} \, \rbf^\dag\cvH{2} \, \cdots
\,\rbf^\dag\cvH{\nSx}\right]^\top$; $\cvH{i}$ is the $i$th column
of the matrix $\ch$; and $\F=\diag{\fv}$. We remark that for a
fixed $\rbf$, the MSE constraint (i.e., the first constraint) at
the optimal solution to problem \eqref{Pb:2:1} must be fulfilled
with equality. We prove it by contradiction. Assume that the MSE
constraint is satisfied with a strict inequality at the optimal
solution $(\av^\star,\{\tbf_i^\star\}_{i=1}^{\nBx})$. By letting
$\bar{\av}=t \av^\star$ for $0<t< 1$, we can choose a sufficient
large $t$ such that
\begin{align}
\hspace{-0.3cm}
        \frac{
                \Norm{{\av^\star}\fv}^2
            }{
                {\av^\star}^\top
                \F
                \Rs
                \F^\dag
                {\av^\star}^*
                +
            \rbf^\dag\Rn\rbf
            }
            >
        \frac{
                \Norm{\bar{\av}^\top\fv}^2
            }{
                \bar{\av}^\top
                \F
                \Rs
                \F^\dag
                \bar{\av}^*
                +
            \rbf^\dag\Rn\rbf
            }
        \geq
        \gamma.
\end{align}
\begin{algorithm}[t]
\caption{ proposed algorithm to solve  $\Pb{2}$}
 \begin{algorithmic}[1]\label{alg:2}
\STATE \textbf{Initialization}: set $n :=0$, and generate  $\av^{(0)}$ and  $\Ampl^{(0)}$.\\[0.1cm]

\STATE \textbf{repeat}\\[0.1cm]

\STATE \quad $\rbf^{(n)}=\left(\ch\Ampl^{(n)}\Rs{\Ampl^{(n)}}^\dag\ch^\dag+\Rn\right)^{-1}\ch\av^{(n)}$\\[0.1cm]

\STATE \quad Solve problem $\sdr{2}$ with $\rbf=\rbf^{(n)}$ to obtain the \\
\quad  optimal value $(\Q^\star,\W^\star)$.\\[0.1cm]

\STATE \quad Set $(\Q^{(n+1)},\W^{(n+1)}):=(\Q^\star,\W^\star)$.\\[0.1cm]

\STATE \quad Construct $\{\av^{(n+1)},\Ampl^{(n+1)}\}$ from $\Q^{(n+1)}$.\\[0.1cm]

\STATE \quad Update $n := n+1$.\\[0.1cm]

\STATE \textbf{until convergence}\\[0.1cm]
\STATE \textbf{Output:}
$(\Q=\Q^{(n)},\W=\W^{(n)},\rbf=\rbf^{(n)})$
\end{algorithmic}
\end{algorithm}
When $\bar{\tbf}_i=t\tbf_i^\star$,
$(\bar{\av},\{\bar{\tbf}_i\}_{i=1}^{\nSx})$ can also be a feasible
solution to problem \eqref{Pb:2:1} with the new objective value
$t^2\sum_{i=1}^{\nBx}\Vnorm{\tbf_i^\star}^2$, which is definitely
smaller than the optimal value when the optimal solution is
$(\av^\star,\{\tbf_i^\star\}_{i=1}^{\nSx})$. This contradicts to
the assumption that $(\av^\star,\{\tbf_i^\star\}_{i=1}^{\nSx})$ is
optimal. Therefore, the MSE constraint must hold with equality.
Let $\Q=\av^* \av^\top$, $\W=\sum_{i=1}^{\nBx}\tbf_i\tbf_i^\dag$,
$\B{\Sigma}=\fv\fv^\top$, $\B{\Psi}=\F\Rs\F^\dag$,
$\G_k=\cvG{k}\cvG{k}^\dag$, and
$\D_k=\mathrm{diag}({0},\ldots,{\Svar^2+\Nvar{k}^2},\ldots,0)$.
 Then, 
as in problem~\eqref{Pb:1:3}, we will omit the rank constraint on
$\Q$ and solve a relaxed version of \eqref{Pb:2:1}, which leads to
\begin{align*}
\begin{aligned}
    &\sdr{2}:\\
    & \underset{\Q,\W}{\text{minimize}}
    &&  \tr\left(\W\right)
\\
    & \text{subject to}
    &&\tr\left(\Q\B{\Sigma}\right)=\gamma\tr\left(\Q\B{\Psi}\right)+\gamma \rbf^\dag\Rn\rbf
\\
    &&& \tr\left(\D_k\Q\right)-\tr\left(\G_k\W\right) \leq 0, \quad \forall k \in \card{\nSx}
\\
    &&& \W \succeq \B{0},  \Q \succeq \B{0}.
\end{aligned}
\end{align*}
The following theorem characterizes the properties of the optimal
solution to problem $\sdr{2}$.
\begin{theorem}[Properties of Optimal Solution]\label{thrm:2}
Let $\beta^\star$ be the dual  optimal solutions associated with
the equality constraint in $\sdr{2}$. We also let $\Q^\star$ and
$\W^\star$ be the primal optimal solutions to $\sdr{2}$. Then the
following three properties are fulfilled:
\begin{enumerate}
\item $\beta^\star>0$; \item
$\rank{\W^\star}\leq\min\left(\nSx,\nRx\right)$; \item
$\rank{\Q^\star}=1$.
\end{enumerate}
\begin{proof}
The proof can be found using the similar steps to the proof for
Theorem~\ref{thrm:1}.
\end{proof}
\end{theorem}

Similarly as in Section~\ref{sec:3}, we  summarize the overall
procedure for solving problem $\Pb{2}$ in Algorithm~\ref{alg:2}.
In this algorithm, the objective value is monotonically reduced in
each step, and for a given feasible threshold $\gamma$, it is
lower-bounded by a certain value. As a result, the algorithm
converges at least to a local optimum. Finally, it can be verified
that the computational complexity of Algorithm~2 is same as that
of Algorithm~\ref{alg:1}.

\subsection{Single-Antenna FC}
It would also be of interest to study problem $\Pb{2}$ for the
single-antenna FC scenario. In this case, problem $\Pb{2}$ can be
rewritten as
\begin{align}\label{Pb:pow:fc}
\begin{aligned}
    & \underset{\av,P}{\text{minimize}}
    & & P
 \\
    & \text{subject to}
    && \frac{
                \av^\top\cvH{}\cvH{}^\dag\av^*
            }{
                \av^\top
                \F
                \Rs
                \F^\dag
                \av^*
                +
                \nvar^2
            }
            =
            \gamma
 \\
    &&& \Norm{\alpha_k}^2\left(\Svar^2+\Nvar{k}^2\right) \leq P \Norm{\RV{g}_k}^2, \,\, \forall k \in \card{\nSx}.
\end{aligned}
\end{align}
Define the matrices $\BB{\Omega}=\left(
  \begin{array}{cc}
   \av^*\av^\top  & \B{0} \\
    \B{0} & P \\
  \end{array}
\right) $; $ \BB{P}=\left(
  \begin{array}{cc}
   \B{0}  & \B{0} \\
    \B{0} & 1 \\
  \end{array}
\right);
$
$ \bar{\D}_k=\left(
  \begin{array}{cc}
    \frac{1}{\Norm{\RV{g}_k}^2}\D_k & \B{0} \\
    \B{0} & -1 \\
  \end{array}
\right); \,\text{and}\,\, \BB{E}=\cvH{}\cvH{}^\dag-\gamma
\F\Rs\F^\dag. $
Then, problem~\eqref{Pb:pow:fc} can be recast as
\begin{align}\label{Pb:pow:fc:1}
\begin{aligned}
    & \underset{\BB{\Omega}}{\text{minimize}}
    & & \tr\left(\BB{\Omega}\BB{P}\right)
 \\
    & \text{subject to}
    &&  \tr\left(\BB{\Omega}\BB{E}\right)
            =
            \gamma\nvar^2
 \\
    &&& \tr\left(\BB{\Omega}\bar{\D}_k\right) \leq 0, \quad \forall k \in \card{\nSx}
\\
    &&& \rank{\BB{\Omega}}=1.
\end{aligned}
\end{align}
By dropping the rank constraint on $\BB{\Omega}$, problem
\eqref{Pb:pow:fc:1} is a SDP and thus can be solved efficiently.
If we denote $\BB{\Omega}^\star$ by the optimal solution  to the
relaxed problem of \eqref{Pb:pow:fc:1}, then
$\rank{\BB{\Omega}^\star}=1$ and the optimal $\av$ and $P$ can be
found from $\BB{\Omega}^\star$. Particularly,
\begin{align}
    P^\star&=\mE{\BB{\Omega}^\star}{\nSx+1}{\nSx+1}\\
    \av^\star &= \sqrt{ \tr(\BB{\Omega}^\star_{\nSx})}\,\uv_1^*,
\end{align}
where $\BB{\Omega}_{\nSx}^\star$  is the $\nSx$th order leading
principal submatrix of $\BB{\Omega}^\star$ obtained by excluding
the $(\nSx+1)$th row and column and $\uv_1$ is the eigenvector
associated with the largest eigenvalue of
$\BB{\Omega}^\star_{\nSx}$. Similarly as in problem $\Pb{1}$, in
this case, the optimal solution $(P^\star, \av^\star)$ is indeed a
global optimum.

\subsection{A Common Energy Harvester}
Now, we consider the converse problem of $\PbS{1}$, in which we
aim to minimize the transmit power at the FC subject to a minimum
requirement of the MSE performance,
\begin{align*}
\begin{aligned}
    \PbS{2}: \quad
    & \underset{\rbf,\av,\tbf}{\text{minimize}}
    & & \Vnorm{\tbf}^2
 \\
    & \text{subject to}
    && \frac{
                \Norm{\rbf^\dag\ch\av}^2
            }{
                \rbf^\dag
                \left(
                    \ch
                    \Ampl
                    \Rs
                    \Ampl^\dag
                    \ch^\dag
                    +
                    \Rn
                \right)
                \rbf
            }
        \geq
            \gamma
\\
    &&& \av^\dag\D \av \leq \Norm{\tbf^\dag \he}^2.
\end{aligned}
\end{align*}
If we multiply $\tbf$ and $\av$ by a scalar $\alpha>1$ and
$\beta<1$, respectively, then the left-hand side of the MSE
constraint (i.e., the first constraint) is strictly increased
while the right-hand side of the sum power constraint (i.e., the
second constraint) as well as the objective function are strictly
decreased. Thus, the optimality for $\PbS{2}$ is achieved when all
the above constraints are satisfied with equality. Problem
$\PbS{2}$ can be formulated as a SDP, and hence solved efficiently
by CVX. In the following, we establish a fundamental relationship
between two problems $\PbS{1}$ and  $\PbS{2}$.

\begin{theorem}[Power--Distortion  Trade-off]\label{thrm:3}
For a distributed estimation system using the BLUE with a common
energy harvester, if we assume that the alternative algorithms
solving $\PbS{1}$ and $\PbS{2}$ are initialized with $\av^{(0)}$,
then the optimal \emph{power--distortion trade-off} is given by
\begin{align}\label{eq:PD}
    \frac{1}{\textsf{mse}}= \fv^\dag \left(\F\Rs\F^\dag+\frac{\rbf^\dag\Rn\rbf}{P \Vnorm{\he}^2}\D\right)^{-1}\fv.
\end{align}
\begin{proof}
See Appendix~\ref{appdx:2}.
\end{proof}
\end{theorem}

Theorem~\ref{thrm:3} is important since it enables to
(numerically) find the \emph{power--distortion trade-off}  for
distributed estimation in the cumulative power constraint case.

\section{Numerical Results}\label{sec:5}
In this section, we provide numerical examples by evaluating our
proposed algorithms in Sections~\ref{sec:3} and \ref{sec:4}.
In the simulations, we consider the widely used 915 MHz frequency band in WSNs \cite{Gut:01:Standard} for both energy and information transmissions. For energy transmission, we consider the use of both the commercially available power transmitter (Powercast TX91501) with transmit power $P=1$W (30 dBm) and the RF power harvester (Powercast P2110). The detailed system parameters are summarized in Table~\ref{table:parameter}. To model a small-scale fading, we assume that the elements of the channel matrices are drawn independently from the Gaussian distribution with zero mean and unit variance. 
To further evaluate the effectiveness of the proposed algorithms,
we also perform comparisons to low-complexity baseline schemes
specified below.

\subsection{Baseline Schemes}
\subsubsection{Suboptimal Design for $\Pb{1}$}
We divide the optimization procedure into two phases. In the first
phase, the energy beamforming vectors $\{\tbf_i\}_{i=1}^{\nBx}$
are designed such that the total harvested energy is maximized,
which leads to the following maximization problem:
\begin{align}\label{Pb:mse:sub:phase1}
\begin{aligned}
    & \underset{\{\tbf_i\}_{i=1}^{\nBx}}{\text{maximize}}
    & & \sum_{k=1}^{\nSx}\beta_k\left(\sum_{i=1}^{\nBx}\Norm{\tbf_i^\dag\cvG{k}}^2\right)
 \\
    & \text{subject to}
    && \sum_{i=1}^{\nBx}\Vnorm{\tbf_i}^2 \leq P.
\end{aligned}
\end{align}
Here, $\{\beta_k\}_{k=1}^{\nSx}$ denote the energy weights
indicating the priority (e.g., sensors with weaker channels can be
assigned to a higher weight to guarantee fairness) of the
corresponding sensors. It has been shown in \cite{ZH:13:WCOM} that
the optimal strategy is to allocate all the power budget to the
direction of $\BB{\eta}$---the eigenvector associated with the
largest eigenvalue of the matrix
$\sum_{k=1}^{\nSx}\beta_k\cvG{k}\cvG{k}^\dag$. The optimal value
in problem \eqref{Pb:mse:sub:phase1} is achieved when
$\tbf_i^\star=\sqrt{p_i} \BB{\eta}$ with $p_i\geq 0$ such that
$\sum_{i=1}^{\nBx}p_i=P$.

In the second phase, we find the amplification vector $\av$ and
the receive filtering $\rbf$ in terms of minimizing the MSE
subject to the energy harvested in the first phase. In particular,
we solve the following problem:
\begin{align}\label{Pb:mse:sub:phase2}
\begin{aligned}
    & \underset{\av,\rbf}{\text{maximize}}
    & & \frac{
                \Norm{\av^\top\fv}^2
            }{
                \av^\top
                \F
                \Rs
                \F^\dag
                \av^*
                +
            \rbf^\dag\Rn\rbf
            }
 \\
    & \text{subject to}
    && \Norm{\alpha_k}^2\left(\Svar^2+\Nvar{k}^2\right) \leq P_k, \quad \forall k \in \card{\nSx},
\end{aligned}
\end{align}
where $\fv=\left[\rbf^\dag\cvH{1} \, \rbf^\dag\cvH{2} \, \cdots
\,\rbf^\dag\cvH{\nSx}\right]^\top$; $\F=\diag{\fv}$; and
$P_k=\Norm{\cvG{k}^\dag\tbf_i^\star}^2$. Problem
\eqref{Pb:mse:sub:phase2} corresponds to problem $\Pb{1}$ without
the total power constraint and can be solved by iteratively
updating $\rbf$ and $\av$.
\subsubsection{Suboptimal Design for $\Pb{2}$}
To reduce the computational burden of the joint optimization for
$\Pb{2}$, we propose a suboptimal design, in which the
optimization procedure is divided into two phases. In the first
phase, we aim to solve the following problem:
\begin{align}\label{Pb:pow:sub:phase1}
\begin{aligned}
    & \underset{\av,\rbf}{\text{minimize}}
    & & \av^\dag\B{D}\,\av
 \\
    & \text{subject to}
    && \frac{
                \Norm{\rbf^\dag\ch\av}^2
            }{
                \rbf^\dag
                \left(
                    \ch
                    \Ampl
                    \Rs
                    \Ampl^\dag
                    \ch^\dag
                    +
                    \Rn
                \right)
                \rbf
            } \geq \gamma.
\end{aligned}
\end{align}
We note that the objective function in problem
\eqref{Pb:pow:sub:phase1} is  the total transmit power of the
sensors. Since the receive filtering $\rbf$  appears only in the
constraint, we can iteratively solve problem
\eqref{Pb:pow:sub:phase1} for $\av$ and $\rbf$. Since the
left-hand side of the constraint is nondecreasing with the norm of
$\av$, the constraint must be satisfied with equality. For a fixed
$\rbf$, the above problem can be expressed as follows:
\begin{align}\label{Pb:pow:sub:phase1:1}
\begin{aligned}
    & \underset{\av,\rbf}{\text{minimize}}
    & & \av^\top\B{D}\,\av^*
 \\
    & \text{subject to}
    && \av^\top\BB{E}\,\av^*=\gamma\,\rbf^\dag\Rn\rbf,
\end{aligned}
\end{align}
where $\BB{E}=\fv\fv^\dag-\gamma \F \Rs \F^\dag$. To guarantee the
feasibility of problem~\eqref{Pb:pow:sub:phase1:1}, the value of
$\gamma$ must be chosen such that $\Norm{\av^\dag\fv}^2 \geq
\gamma\, \av^\top\F \Rs \F^\dag\,\av^*$. Since the quantities
$\av^\top\B{D}\,\av^* \geq 0$ and $\av^\top\BB{E}\,\av^* \geq 0$
are positive, problem \eqref{Pb:pow:sub:phase1:1} can be rewritten
as
\begin{align}\label{Pb:pow:sub:phase1:2}
\begin{aligned}
    & \underset{\av,\rbf}{\text{maximize}}
    & & \frac{\av^\top\BB{E}\,\av^*}{\av^\top\B{D}\,\av^*}
 \\
    & \text{subject to}
    && \av^\top\BB{E}\,\av^*=\gamma\,\rbf^\dag\Rn\rbf
\end{aligned},
\end{align}
which is a Rayleigh quotient. Thus, the optimal solution to
problem \eqref{Pb:pow:sub:phase1:2} is given by
\begin{align}
    \av^\star
    =
        \sqrt{\frac{\gamma\,\rbf^\dag\Rn\rbf}{\uv_1^\dag \B{D}^{-1/2} \BB{E} \B{D}^{-1/2}\uv_1}}
        \B{D}^{-1/2}
        \uv_1^*,
\end{align}
where $\uv_1$ denotes the unit-norm eigenvector associated with
the largest eigenvalue of the matrix $\B{D}^{-1/2} \BB{E}
\B{D}^{-1/2}$, $\maxeig\left(\B{D}^{-1/2} \BB{E}
\B{D}^{-1/2}\right)$. It follows that the minimum total transmit
power of the sensors, $P_\textsf{s}^\star$, required to achieve
the MSE of $1/\gamma$ is given by
\begin{align}
    P_\textsf{s}^\star=\frac{\gamma\,\rbf^\dag\Rn\rbf}{\maxeig\left( \B{D}^{-1/2} \BB{E} \B{D}^{-1/2}\right)}.
\end{align}

In the second phase, we aim to minimize the total transmit power
at the FC with the amplification coefficients
$\{\alpha_k\}_{k=1}^{\nSx}$ that are the solutions to problem
\eqref{Pb:pow:sub:phase1}. In other words, we  find the optimal
solution to the following minimization problem:
\begin{align}\label{Pb:pow:sub:phase2}
\begin{aligned}
    & \hspace{-0.3cm} \underset{\{\tbf_i\}_{i=1}^{\nBx}}{\text{minimize}} \hspace{-0.2cm}
    & & \sum_{i=1}^{\nBx}\Vnorm{\tbf_i}^2
 \\
    &\hspace{-0.3cm} \text{subject to}  \hspace{-0.2cm}
    && \sum_{i=1}^{\nBx}\Norm{\tbf_i^\dag\cvG{k}}^2  \geq \Norm{\alpha_k}^2\left(\Svar^2+\Nvar{k}^2\right) , \, \forall k \in \card{\nSx}.
\end{aligned}
\end{align}
Problem \eqref{Pb:pow:sub:phase2} can be effectively solved by
CVX. In the following subsections, we use these suboptimal designs
as the baseline schemes to assess the effectiveness of our
proposed algorithms.
\begin{table}[t!]
    \caption{
                System Parameters
    }
    \label{table:parameter}
\begin{center}
 \begin{tabular}{c|c}
        \hline
        Parameter & Value
        \bigstrut \\ \hline \hline
        Network Topology & {\centering $10 \textsf{m} \times 10 \textsf{m}$ square box}  \bigstrut\\
         \hline
         &
        Located at the origin $\left(0,0\right)$\bigstrut\\
        \cline{2-2}
        & Transmission power $30$ dBm
       \bigstrut\\
       \cline{2-2}
       \raisebox{0.0cm}[0.0cm][0.0cm]{Fusion Center}
        &  \multirow{3}{5cm}{
        \centering  \vspace{0.25cm}  Receiver noise power $-103.16$ dBm } \\[0.2cm]
        & (effective noise bandwidth $2$ MHz   \\[-0.1cm]
        & and noise figure 7 dB)
        \bigstrut\\
        \hline
       \multirow{2}{1.5cm}{\centering Sensors } &
       Placed uniformly over $\{x,y| x,y \in \left[-10,10\right]\}$
        \bigstrut\\
        \cline{2-2}
        &
        Energy harvesting efficiency $51\%$
        \bigstrut\\    \hline
        Path loss \cite{SR:92:AP} &
        $\PL{d}=31.7 +27.6 \log_{10}\left(d_{[\textsf{meters}]}\right) \quad [\textsf{dB}]$
        \bigstrut\\
        \hline
    \end{tabular}
\end{center}
\end{table}

\subsection{MSE Minimization}

\begin{figure}[t!]
\centerline{\includegraphics[width=0.52\textwidth]{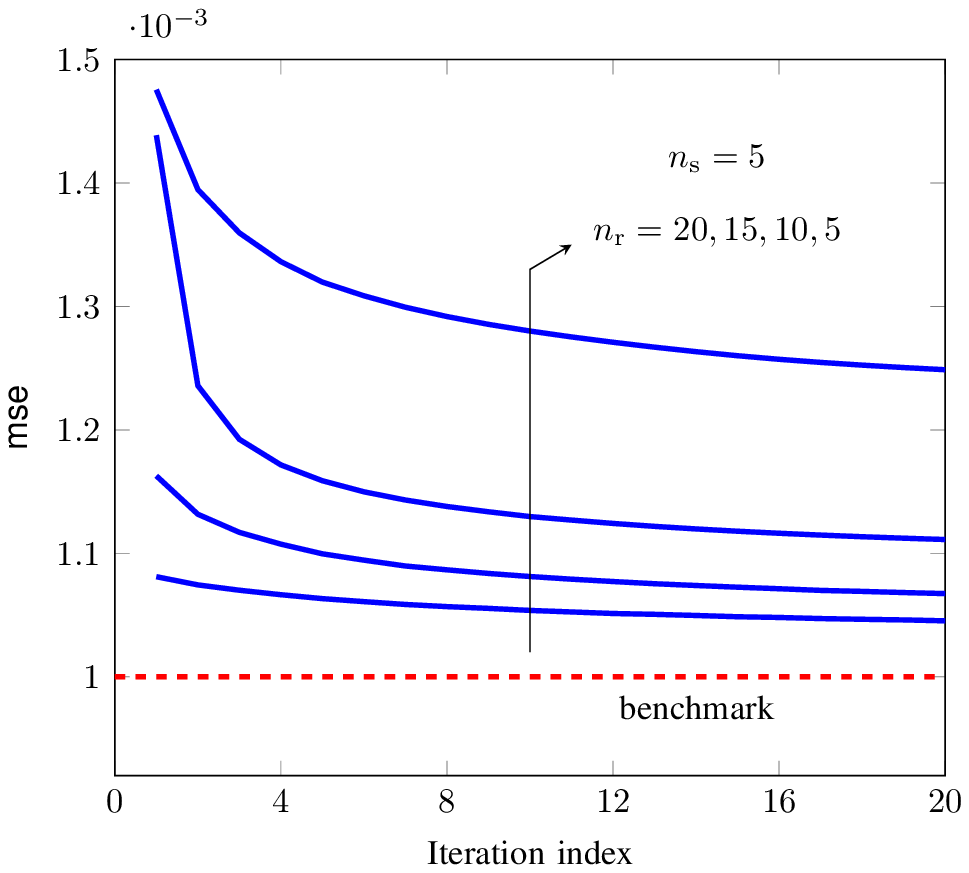}}
\caption{The average MSE for distributed estimation versus
iteration index when $\nSx=5$, $\Rs=0.1\IM{\nSx}$, $P=30$ dBm,
$\Svar=1$, and $\nRx=5,10,15,20$. } \label{fig:1}
\end{figure}

\begin{figure}[t]
\centerline{\includegraphics[width=0.52\textwidth]{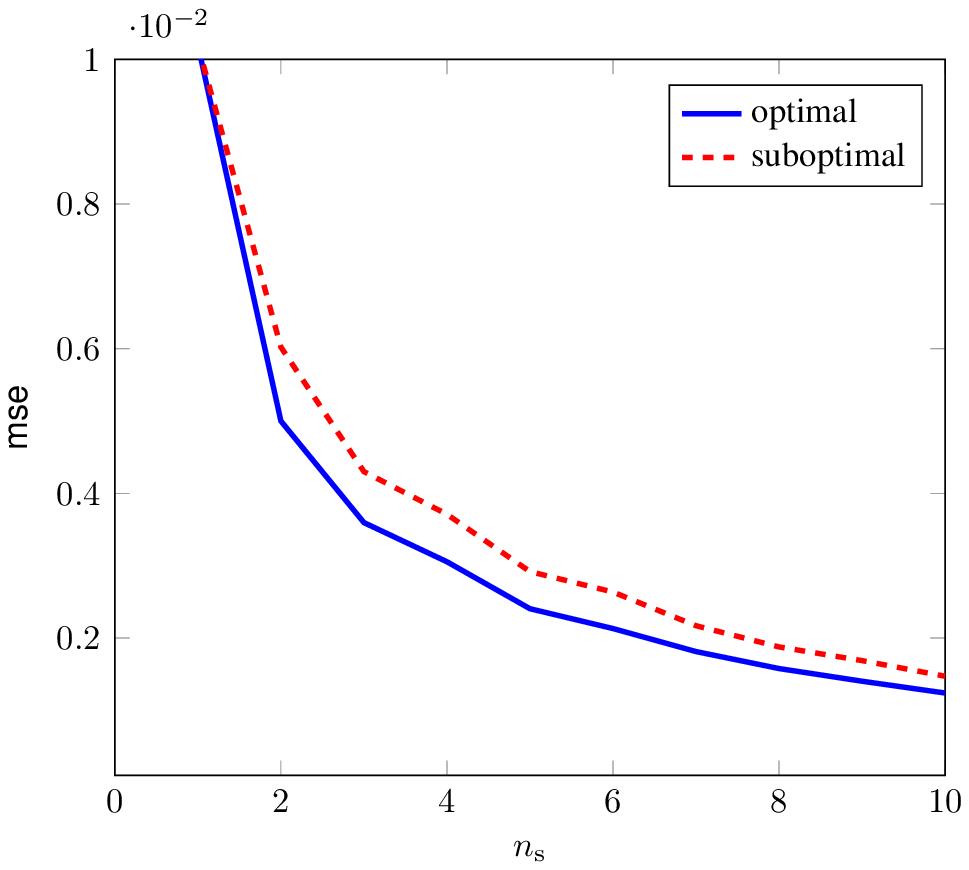}}
\caption{The average MSE for distributed estimation as a function
of $\nSx$ when $\nRx=5$, $P=30$ dBm, $\Rs=0.1\IM{\nSx}$, and
$\Svar=1$.} \label{fig:2}
\end{figure}

Figure~\ref{fig:1} shows the average MSE for distributed
estimation versus iteration index when $\nSx=5$,
$\Rs=10^{-2}\IM{\nSx}$,  $P=30$ dBm, $\Svar=1$, and
$\nRx=5,10,15,20$. One can see that the average MSE monotonically
decreases while the algorithm converges within a few iterations.
It can be obviously seen that the MSE performance is improved with
the increasing number of antennas at the FC, $\nRx$. In this
figure, we also plot a benchmark ideal case for distributed
estimation, where all the observations at the sensors are assumed
to be directly available at the FC, which will give a lower bound
on the MSE performance. One can see that the average MSE evaluated
via our simulation tends to approach the benchmark value,
$\left[\B{1}^\dag\Rs^{-1}\B{1}\right]^{-1}$, as $\nRx$ increases.
In Fig.~\ref{fig:2}, the average MSE for distributed estimation is
shown as a function of $\nSx$ for the optimal and suboptimal
solutions when $P=30$ dBm, $\nRx=5$, $\Rs=0.1\IM{\nSx}$,
$\Rn=0.5\IM{\nRx}$, and $\Svar=1$. As expected, the MSE
performance is improved as $\nSx$ increases. In this example, we
can see that the suboptimal solution shows a reasonable
performance compared to the optimal one.

\begin{table}[t]
    \caption{
                Power Control
    }
    \label{table:1}
\begin{center}
    \begin{tabular}{c|c|c}
        \hline \hline
        Sensor index & Harvested power [dBm] & Transmit power [dBm]
         \bigstrut\\
        \hline
        1 &  -31.449  &   -31.449  \bigstrut\\
        2 &  -27.687    &    -34.347 \\[0.3ex]
        3 &  -30.737    &    -30.737 \\[0.3ex]
        4 &  -32.865    &    -32.865 \\[0.3ex]
        5 &  -13.847    &    -48.067 \\[0.3ex]
        6 &  -29.886    &    -31.999\\[0.3ex]
        7 &  -28.307    &    -32.964 \\[0.3ex]
        \hline
    \end{tabular}
\end{center}
\end{table}
\begin{figure}[t]
\centerline{\includegraphics[width=0.52\textwidth]{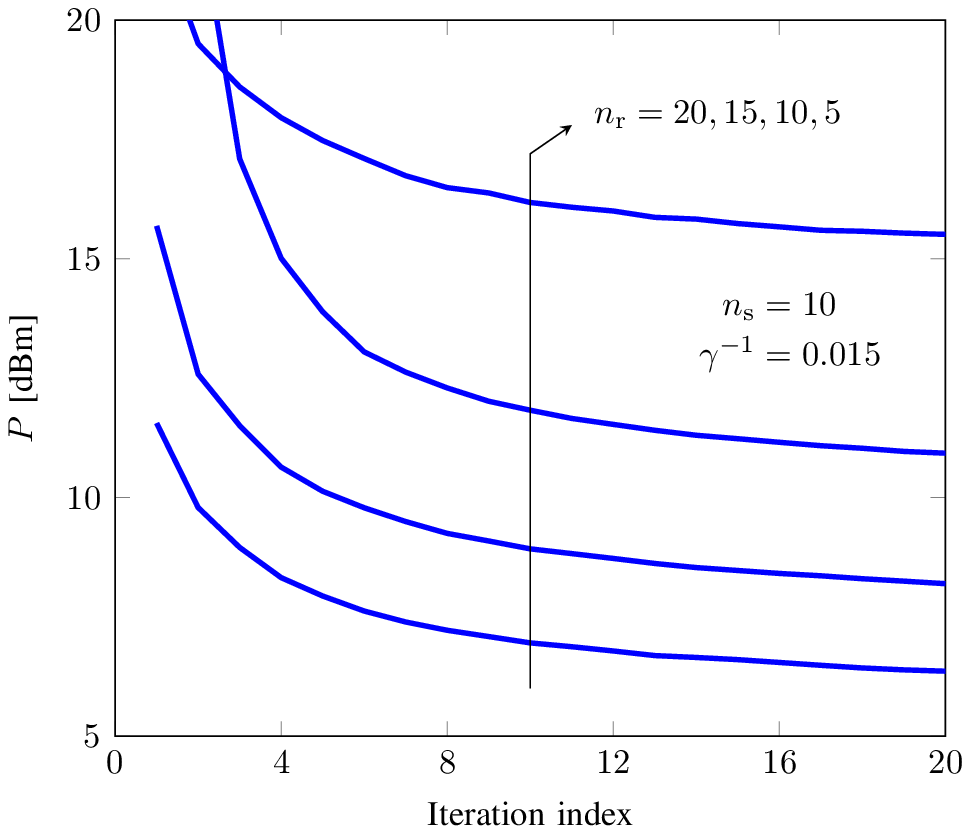}}
\caption{The average minimum  transmit power at the FC for
distributed estimation  versus iteration index when $\nSx=10$,
$\Rs=0.1\IM{\nSx}$,  $\gamma^{-1}=0.015$, $\Svar=1$, and
$\nRx=5,10,15,20$. } \label{fig:4}
\end{figure}

\begin{figure}[t]
\centerline{\includegraphics[width=0.52\textwidth]{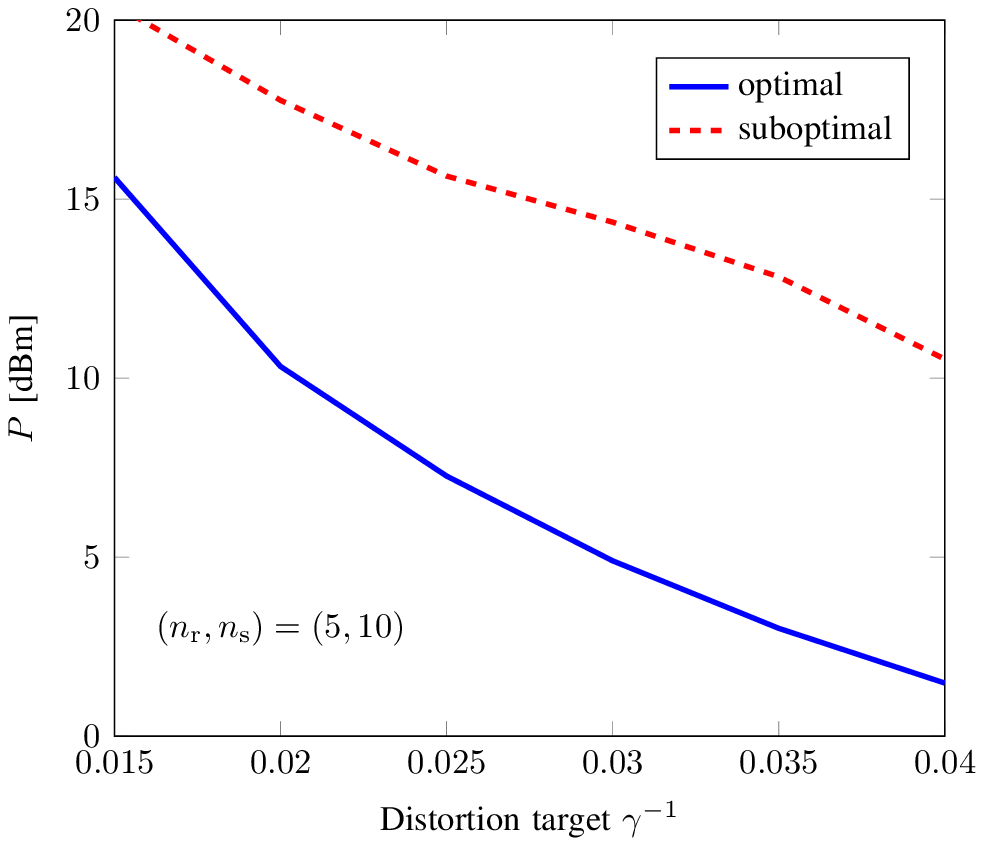}}
\caption{The average minimum transmit power at the FC for
distributed estimation as a function of the distortion target
$\gamma^{-1}$ when $\nSx=10$, $\Rs=0.1\IM{\nSx}$, $\Svar=1$, and
$\nRx=5$. } \label{fig:5}
\end{figure}


In Table~\ref{table:1}, in order to elaborate on the attributes of
the optimal solution to the MSE minimization problem, we present
the values of the harvested and transmit power of each sensor at
the optimal solution to problem $\Pb{1}$. In this example, we set
$P=30$ dBm, $\Rs=0.1\IM{\nSx}$, $\nSx=7$, and $\nRx=2$.  One can
see that some of the sensors do not use their maximum power
harvested from the FC, which implies that power control is needed
to guarantee the optimal solution. In other words, some of
individual power constraints (i.e., the first constraint) in
problem $\Pb{1}$ may not be fully utilized, or equivalently, the
corresponding dual variables may be zero. This is attributed from
the fact that transmission with the full power may increase the
interference level at the FC, which in turn reduces the estimation
reliability. In this example, sensors 2, 5, 6, and 7 use only  a
fraction of their harvested power.
\begin{figure}[t]
\centerline{\includegraphics[width=0.52\textwidth]{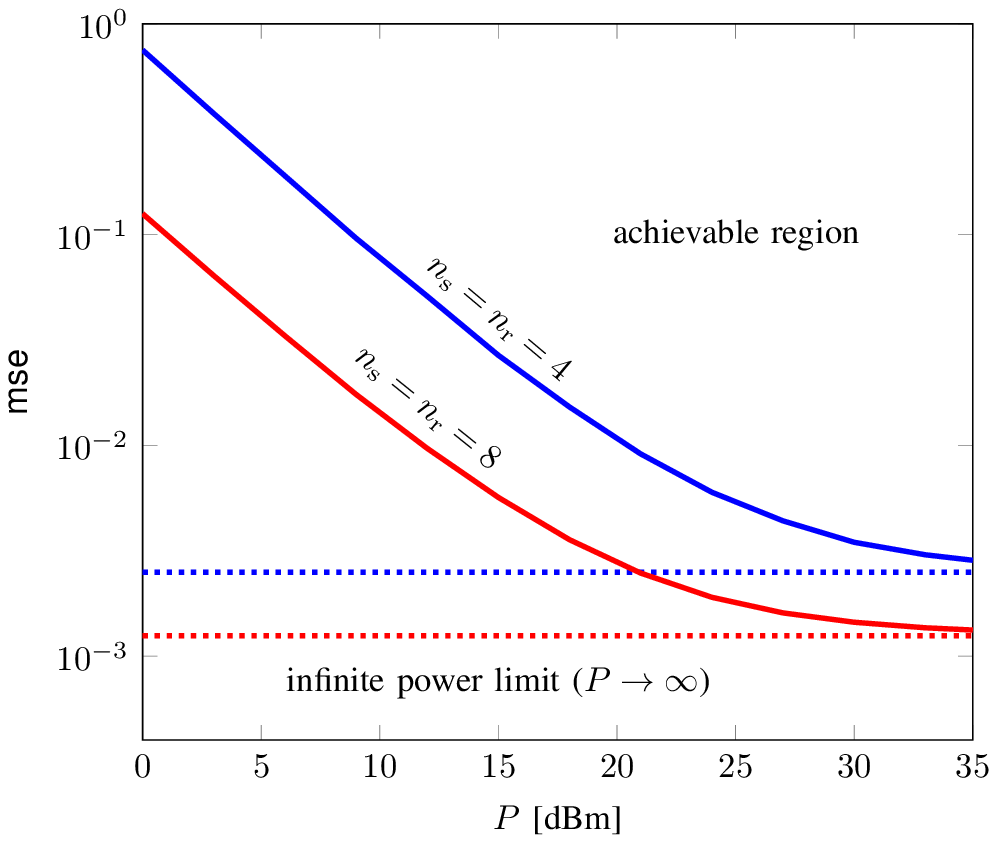}}
\caption{The distortion--power trade-off for distributed
estimation with a common energy harvester when
$\Rs=10^{-2}\IM{\nSx}$, $\Svar=1$, $\nSx=\nRx=4$, and
$\nSx=\nRx=8$. } \label{fig:6}
\end{figure}
\subsection{Total Power Minimization}
Figure~\ref{fig:4} illustrates the average minimum  transmit power
at the FC for distributed estimation versus iteration index at the
distortion target of $\gamma^{-1}=0.015$ when $\nSx=10$,
$\Rs=0.1\IM{\nSx}$, $\Svar=1$, and $\nRx=5,10,15,20$. As shown in
this figure,  the proposed algorithm converges quickly, and the
transmit power is reduced as the number of antennas, $\nRx$,
increases. In Fig.~\ref{fig:5},  the average minimum  transmit
power at the FC for distributed estimation is shown as a function
of the distortion target $\gamma^{-1}$ when $\nSx=10$,
$\Rs=0.1\IM{\nSx}$, $\Svar=1$, and $\nRx=5$. It is clear that the
more strict distortion requirement is, the more power is needed.
Note that the distortion target must be no smaller than the
benchmark MSE value such that the optimization problem is
feasible. One can also see that the optimal scheme should be used
for power saving.
 In this example, we can save the amount of transmit power of 7.44,  9.46, and 9.05 dBm at $\gamma^{-1}=0.02, 0.03, 0.04$, respectively, compared to the suboptimal case.

\subsection{A Common Energy Harvester}
Finally, we validate the performance of the distributed estimation
system with a common energy harvester. Specifically, the
power--distortion trade-off is ascertained by referring to
Fig.~\ref{fig:6}, where the optimal MSE is depicted as a function
of the minimum transmit power $P$ for distributed estimation when
$\Rs=10^{-2}\IM{\nSx}$, $\Svar=1$, $\nSx=\nRx=4$, and
$\nSx=\nRx=8$.  In this figure,  the region above each trade-off
curve is achievable. As $P$ tends to infinity, the MSE converges
to that of centralized estimations, i.e.,
$\left[\B{1}^\dag\Rs^{-1}\B{1}\right]^{-1}$, plotted with the
dotted curve.
Moreover, as expected, the achievable region gets broader for a
larger $(\nSx,\nRx)$ pair.

\section{Concluding Remarks}\label{sec:6}
Using the SDR, we developed a new framework for solving the
network lifetime problem of a WSN. To that end, we adopted the
notion of RF-based WPT as well as the multiple-antenna technology
so that both the life span and the estimation performance are
substantially improved. In this paper, two optimization problems
were formulated and iteratively  solved by two proposed
algorithms, which turned out to  guarantee the convergence at
least to a local optimum. We showed that power control is indeed
required at the optimal solution. It was also shown that having
multiple antennas at the FC provides a significant improvement in
the estimation performance. Especially, it was shown that  as the
number of antennas grows large, the MSE of the distributed
estimation with the BLUE approaches that of centralized
estimations. \appendices
\section{Proof of Theorem~\ref{thrm:1}}\label{appdx:1}
\begin{proof}
We start by proving the first property of Theorem~\ref{thrm:1}. We
exploit the strong duality  and then examine the
Karush-Kuhn-Tucker condition of $\sdr{1}$. Let $\nu$,
$\{\lambda_k\}_{k=1}^{\nSx}$, and $\beta$ be the dual variables of
problem $\sdr{1}$. The Lagrangian of problem $\sdr{1}$ is defined
as
\begin{align*}
    &\mathcal{L}\left(\nQ,\nW,\eta,\nu,{\lambda_k},\beta\right)
    =
        -\tr\left(\nQ\B{\Sigma}\right)
        +
        \beta\left(\tr\left(\nW\right)-\eta P\right)
    \nonumber\\
    &\hspace{-0.2cm}
        +
        \sum_{k=1}^{\nSx}
            \lambda_k
            \tr\left(
                \D_k
                \nQ
                -
                \G_k
                \nW
            \right)
        +
        \nu\left(
            \tr\left(\nQ\B{\Psi}\right)
            +
            \eta\rbf^\dag\Rn\rbf
            -
            1
        \right).
\end{align*}
Then, the dual function of problem $\sdr{1}$ is given by
\begin{align*}
\min_{\nQ \succeq \B{0},\nW \succeq \B{0},\eta>0}
\mathcal{L}\left(\nQ,\nW,\eta,\nu,{\lambda_k},\beta\right),
\end{align*}
which can be equivalently expressed as
\begin{align}\label{eq:lgr}
    \min_{\nQ \succeq \B{0},\nW \succeq \B{0},\eta>0}
        \tr\left(\nQ\Y\right)
        +
        \tr\left(\nW\Z\right)
        +
        \eta\xi
        -
        \nu,
\end{align}
where
\begin{align*}
    \xi&=\nu\rbf^\dag\Rn\rbf-\beta P\\
    \Y&=-\B{\Sigma}+\sum_{k=1}^{\nSx}\lambda_k\D_k+\nu\B{\Psi}\\
    \Z&=-\sum_{k=1}^{\nSx}\lambda_k\G_k+\beta\IM{}.
\end{align*}
When we let $\nu^\star$, $\{\lambda_k^\star\}_{k=1}^{\nSx}$, and
$\beta^\star$ be the optimal dual solutions to problem $\sdr{1}$,
we define
\begin{align}\label{eq:Y}
\Y^\star&=-\B{\Sigma}+\sum_{k=1}^{\nSx}\lambda_k^\star\D_k+\nu^\star\B{\Psi}\\
    \label{eq:Z}
    \Z^\star&=-\sum_{k=1}^{\nSx}\lambda_k^\star\G_k+\beta^\star\IM{}.
\end{align}
Then, the optimal $\nQ^\star$ must be the solution to the
following problem:
\begin{align}\label{Pb:nQ}
\begin{aligned}
    & \underset{\nQ \succeq \B{0}}{\text{minimize}}
    &&  \tr\left(\nQ\Y^\star\right).
\end{aligned}
\end{align}
To guarantee a bounded optimal value, we must have $\Y^\star
\succeq \B{0}$, and hence we obtain the optimal value
$\tr(\nQ^\star\Y^\star)=0$, which  implies that
\begin{align}\label{eq:Y:cond}
\nQ^\star\Y^\star=\B{0}.
\end{align}

In the same manner, it follows that $\nW^\star\Z^\star=\B{0}$ and
$\eta^\star \xi^\star=0$, or equivalently $\xi^\star=0$ since
$\eta^\star >0$, where
$\xi^\star=\nu^\star\rbf^\dag\Rn\rbf-\beta^\star P$. From
\eqref{eq:lgr}, the dual problem to  problem $\sdr{1}$ can be
rewritten as follows:
\begin{align}\label{Pb:dual}
\begin{aligned}
    & \underset{\nu,\{\lambda_k\}_{k=1}^{\nSx},\beta}{\text{minimize}}
    &&  \nu
\\
    & \text{subject to}
    &&  \Y \succeq \B{0}, \Z \succeq \B{0}, \xi \geq 0
\\
    &&& \beta \geq 0, \lambda_k \geq 0, \quad \forall k \in \card{\nSx}.
\end{aligned}
\end{align}
Since the duality gap between  problem $\sdr{1}$ and
\eqref{Pb:dual} is zero, $\nu^\star$ is equal to the optimal value
of problem $\sdr{1}$, which is positive. Thus, we conclude that
$\nu^\star>0$. Next, we will show that $\beta^\star >0$. First, if
there exits a $\lambda_k>0$, then from the condition $\Z \succeq
\B{0}$, it follows that $\beta^\star >0$. From the condition
$\xi^\star=0$ and the facts that $\Rn \succ \B{0}$ and $\nu^\star
>0$, we also conclude that $\beta^\star >0$.

To prove the second property of Theorem~\ref{thrm:1}, we use the
fact that for any two matrices of the same size $\B{A}$ and
$\B{B}$,  $\rank{\B{A}-\B{B}}\geq
\Norm{\rank{\B{A}}-\rank{\B{B}}}$ \cite{HJ:85:Book}. Since
$\beta^\star>0$ and
$\rank{\sum_{k=1}^{\nSx}\lambda_k^\star\G_k}\leq \nSx$, it follows
from \eqref{eq:Z} that $\rank{\Z^\star} \geq \Norm{\nRx-\nSx} \geq
\nRx-\nSx$. Let $\textsf{Null}\left(\Z^\star\right)$ be the null
space of $\Z^\star$. Then from the condition
$\nW^\star\Z^\star=\B{0}$, we must have $\nW^\star \in
\textsf{Null}\left(\Z^\star\right)$. Since  $\rank{\Z^\star} \geq
\nRx-\nSx$ and $\rank{\nW^\star} \leq
\textsf{dim}\left(\textsf{Null}\left(\Z^\star\right)\right)$, it
follows that $\rank{\nW^\star}\leq \nSx$. Using the fact that
$\nW^\star$ is an $\nRx \times \nRx$ matrix, we conclude that
$\rank{\nW^\star}\leq \nRx$.

Finally, we prove the property of the optimal solution
$\nQ^\star$. Since $\B{\Psi} \succ \B{0}$,
$\sum_{k=1}^{\nSx}\lambda_k^\star\D_k \succeq \B{0}$, and
$\nu^\star > 0$, we obtain
\begin{align}
    \rank{\nu^\star\B{\Psi} + \sum_{k=1}^{\nSx}\lambda_k^\star\D_k}=\nSx.
\end{align}
Hence, from the definition of $\Y^\star$ in \eqref{eq:Y}, it
follows that
\begin{align}\label{eq:Y:rank}
    \rank{\Y^\star} \geq \nSx -\rank{\B{\Sigma}}=\nSx-1.
\end{align}
From the condition \eqref{eq:Y:cond}, $\nQ^\star$ must lie in the
null space of $\Y^\star$. Therefore, $\rank{\nQ^\star} \leq
\textsf{dim}\left(\textsf{Null}\left(\Y^\star\right)\right)$,
which is upper-bounded by one due to \eqref{eq:Y:rank}. Now,
assume that $\rank{\Y^\star}=\nSx$. Then from \eqref{eq:Y:cond},
it follows that $\nQ^\star=\B{0}$, which cannot be the optimal
solution to  problem $\sdr{1}$. In consequence, we must have
$\rank{\Y^\star}=\nSx-1$ and thus $\rank{\nQ^\star}=1$, which
completes the proof of Theorem~\ref{thrm:1}.
\end{proof}
\section{Proof of Theorem~\ref{thrm:3}}\label{appdx:2}
\begin{proof}
We derive the result in \eqref{eq:PD} by showing that the optimal
MSE in $\PbS{1}$ and the optimal power in $\PbS{2}$ are the
inverse of each other. We start the proof by introducing the
following lemma.
\begin{lemma}\label{lem:PD}
For a given $\rbf$ in \eqref{eq:rbf}, let
$(\av_1^\star,\tbf_1^\star)$ and $(\av_2^\star,\tbf_2^\star)$ be
the optimal solutions to problems $\PbS{1}$ and $\PbS{2}$,
respectively. We also let $\f{1}{\cdot,\cdot}$ and
$\f{2}{\cdot,\cdot}$ be the objective functions in $\PbS{1}$ and
$\PbS{2}$, respectively. Then, they obey the property: if
$\gamma=\f{1}{\av_1^\star,\tbf_1^\star}$, then
$(\av_2^\star,\tbf_2^\star):=(\av_1^\star,\tbf_1^\star)$; and if
$P=\f{2}{\av_2^\star,\tbf_2^\star} $, then
$(\av_1^\star,\tbf_1^\star):=(\av_2^\star,\tbf_2^\star)$.
\begin{proof}
First, it is worth noting that all the inequality constraints in
$\PbS{1}$ and $\PbS{2}$ are satisfied with equality at the optimal
solutions. For a given $\rbf$, we have $\Vnorm{\tbf_1^\star}^2=P$.
We will prove that $(\av_1^\star,\tbf_1^\star)$ is also a solution
to $\PbS{2}$, i.e.,
$(\av_2^\star,\tbf_2^\star):=(\av_1^\star,\tbf_1^\star)$. We prove
it by contradiction. Assume that $(\av_1^\star,\tbf_1^\star)$ is
not a solution to $\PbS{2}$, that is, there exists a feasible
solution $(\av_1',\tbf_1')$ to $\PbS{2}$ such that
$\Vnorm{\tbf_1'} < \Vnorm{\tbf_1^\star}$. In other words, we can
find a constant $c>1$ such that
\begin{align}\label{eq:lem:1}
    \Vnorm{\tbf_1'} < c\Vnorm{\tbf_1'} \leq \Vnorm{\tbf_1^\star}.
\end{align}
Since the objective function in $\PbS{1}$ is monotonically
increasing with the norm of $\av$, it follows that $\f{1}{c\,
\av_1',c \,\tbf_1'} > \f{1}{\av_1',\tbf_1'}$. Since
$(\av_1',\tbf_1')$ is feasible to $\PbS{1}$, we also have that $
\f{1}{\av_1',\tbf_1'}\geq \f{1}{\av_1^\star,\tbf_1^\star}$. Thus,
we obtain
\begin{align}
     \f{1}{c\, \av_1',c\, \tbf_1'}
     >
     \f{1}{\av_1',\tbf_1'}
     \geq
     \f{1}{\av_1^\star,\tbf_1^\star}.
\end{align}
From \eqref{eq:lem:1} and the second constraint in $\PbS{1}$, the
solution $(c\, \av_1',c\, \tbf_1')$ is also feasible to $\PbS{1}$,
and yields a higher objective value than the optimal
$(\av_1^\star,\tbf_1^\star)$ does. This contradicts to the
assumption that $(\av_1^\star,\tbf_1^\star)$ is optimal to
$\PbS{1}$, and thus  $(\av_1^\star,\tbf_1^\star)$  must be a
solution to $\Pb{2}$.

The proof for the second claim can be found using the similar
steps to the proof for the first one, and hence is omitted here.
\end{proof}
\end{lemma}
From \eqref{eq:mse:sum:obj}, the optimal $\textsf{mse}$ satisfies
\eqref{eq:PD}. Since the optimal $P$ and  $\textsf{mse}$ are the
inverse of each other, it follows that optimal $P$ also satisfies
\eqref{eq:mse:sum:obj}. Due to the fact that the above property
holds for any $\rbf$, it holds for the optimal $\rbf$ as well,
which completes the proof of Theorem~\ref{thrm:3}.
\end{proof}


\end{document}